\documentclass[preprint,floatfix,showpacs,aps,nofootinbib]{revtex4}
\usepackage{graphicx}
\usepackage{dcolumn}
\usepackage{bm}
\usepackage{amsmath,amssymb}
\usepackage{bbold}

\newcommand {\be}      {\begin{equation}}   
\newcommand {\ee}      {\end{equation}}
\newcommand {\bee}     {\begin{equation*}}   
\newcommand {\eee}     {\end{equation*}}
\newcommand {\bea}     {\begin{eqnarray}}   
\newcommand {\eea}     {\end{eqnarray}}
\newcommand {\beaa}    {\begin{eqnarray*}}   
\newcommand {\eeaa}    {\end{eqnarray*}}
\newcommand {\bse}     {\begin{subequations}}
\newcommand {\ese}     {\end{subequations}}
\newcommand {\nn}      {\nonumber}
\newcommand {\lam}     {\lambda}
\newcommand {\sig}     {\sigma}
\newcommand {\del}     {\delta}
\newcommand {\eps}     {\epsilon}
\newcommand {\pd}      {\partial}
\newcommand {\pdi}     {\pd_i}
\newcommand {\pdj}     {\pd_j}
\newcommand {\piai}    {\pi_{A_i}}
\newcommand {\pj}      {\pd_j}
\newcommand {\pk}      {\pd_k}
\newcommand {\pt}      {\pd_t}
\newcommand {\pr}      {\prime}
\newcommand {\cu}      {{\cal U}}
\newcommand {\cl}      {{\cal L}}
\newcommand {\ch}      {{\cal H}}
\newcommand {\cv}      {{\cal V}}
\newcommand {\cs}      {{\cal S}}
\newcommand {\ceta}    {{\cal N}}
\newcommand {\chns}    {\ch_{NS}}
\newcommand {\hns}     {H_{NS}}
\newcommand {\che}     {\ch_E}
\newcommand {\cle}     {\cl_E}
\newcommand {\clns}    {\cl_{NS}}
\newcommand {\du}      {d\cu}
\newcommand {\coe}     {{\cal O}_E}
\newcommand {\com}     {{\cal O}_m}
\newcommand {\cons}    {{\cal O}_{NS}}
\newcommand {\intx}    {\int d^3x}

\newcommand {\onehalf} {\frac{1}{2}}
\newcommand {\vect}[1] {\mathbf{#1}}
\newcommand {\bx}      {\vect{x}}
\newcommand {\ba}      {\vect{A}}
\newcommand {\bb}      {\vect{\Pi}}
\newcommand {\bax}     {\ba(\bx,t)}
\newcommand {\bv}      {\vect{v}}
\newcommand {\bq}      {\vect{q}}

\newcommand {\bnh}     {\hat{\vect{n}}}
\newcommand {\bn}      {\vect{\nabla}}

\newcommand {\crr}      {{\cal\vect{R}}_{cm}}

\begin{document}

\title{Navier-Stokes Hamiltonian for the \linebreak Similarity Renormalization Group}

\author{Billy D. Jones} \email{bdjwww@uw.edu}
\affiliation{Applied Physics Laboratory, University of Washington, Seattle, WA 98105}

\begin{abstract} 

The Navier-Stokes Hamiltonian is derived from first principles. Its 
Hamilton equations are shown to be equivalent to the continuity, 
Navier-Stokes, and energy conservation equations of a compressible viscous 
fluid. The derivations of the Euler and Navier-Stokes Hamiltonians 
are compared, with the former having identical dynamics to the 
Euler equation with the viscosity terms of the Navier-Stokes 
equation dropped from the beginning. The two Hamiltonians have the 
same number of degrees of freedom in three spatial and 
one temporal dimension: six independent scalar potentials, but their dynamical 
fields are necessarily different due to a theory with dissipation 
not mapping smoothly onto one without. The dynamical coordinate field 
of a dissipative fluid is a vector field that stores 
the initial position of all of its fluid particles. Thus these 
appear to be natural coordinates for studying arbitrary separations of 
fluid particles over time. The classical similarity renormalization group is 
introduced and the first steps are carried out deriving a 
flow equation for the Navier-Stokes Hamiltonian. Finally, the symmetries 
of a nonrelativistic viscous fluid are discussed through its galilean 
algebra with dissipative canonical Poisson brackets.

\end{abstract}


\maketitle
\thispagestyle{empty}

\tableofcontents
\thispagestyle{empty}
\newpage

\setcounter{page}{1}


\section{Introduction}

The Navier-Stokes Hamiltonian ($\hns$) is derived from first principles. 
The goal is to come up with useful dynamical degrees of freedom 
in order to systematically integrate the equations 
of motion of a fluid one scale at a time with the 
similarity renormalization group (SRG) using convenient approximations set up for a Hamiltonian. 
These hamiltonian renormalization group studies are saved for later work, however a flow equation 
for $\hns$ is derived in Appendix~\ref{app:csrg}. This paper 
motivates the program and derives $\hns$ from first principles along with 
its dissipative canonical Poisson bracket with a general classical dissipative observable. 
First, we back up to the beginning in order to better know where $\hns$ truly came from.

We start in a nonrelativistic framework (with all fluid particle 
speeds of interest much less than the speed of light) and further assume that the 
mass density (hereafter just ``density'') is high enough (and/or collisions strong enough) such that the continuum 
approximation is valid, yet low enough (and/or temperature high enough) such that quantum effects are negligible. 
This implies that the mean free path is very small with respect to the smallest scale of turbulence, 
the Kolmogorov or dissipation scale (see Fig.~1) \cite{sm}:
\bee
\ell\sim\frac{1}{n\,\sig_{tot}}\ll \eta_{_{Kol}}~~~[{\rm then}~{\rm continuum}~{\rm approx.}~{\rm valid}]\;,
\eee
where $\ell$ is the mean free path, 
$n$ is the number density, and $\sig_{tot}$ is the total molecular cross section. 
It also implies that the chemical potential $\mu\rightarrow-\infty$ as holds for classical physics. 
Then, for example, for a classical ideal gas we have 
\bee
n\lam^3_{_T}=e^{\frac{\mu}{k_BT}}~\stackrel{^{\mu\rightarrow-\infty}}{\ll}
~1~~~[{\rm then}~{\rm quantum}~{\rm effects}~{\rm negligible}]\;,
\eee
where $k_B$ is Boltzmann's constant, T is temperature, and $\lam_{_T}$ is the thermal wavelength: 
\bee
\lam_{_T}\sim\frac{2\pi\hbar}{p_{_T}}\sim\frac{2\pi\hbar}{\sqrt{2\pi mk_BT}}\;,
\eee
with $p_{_T}$ the thermal momentum, $m$ the mass of a single molecule of the fluid, 
and $\hbar$ the quantum of action (which makes this quantity 
tiny except for very small temperatures which we assume do not occur). 
In summary: we start with a classical nonrelativistic fluid (e.g.\ air or water under standard 
ambient conditions). 

\vspace{12pt}
\begin{figure}[!ht]
\centering
    \includegraphics[scale=0.75]{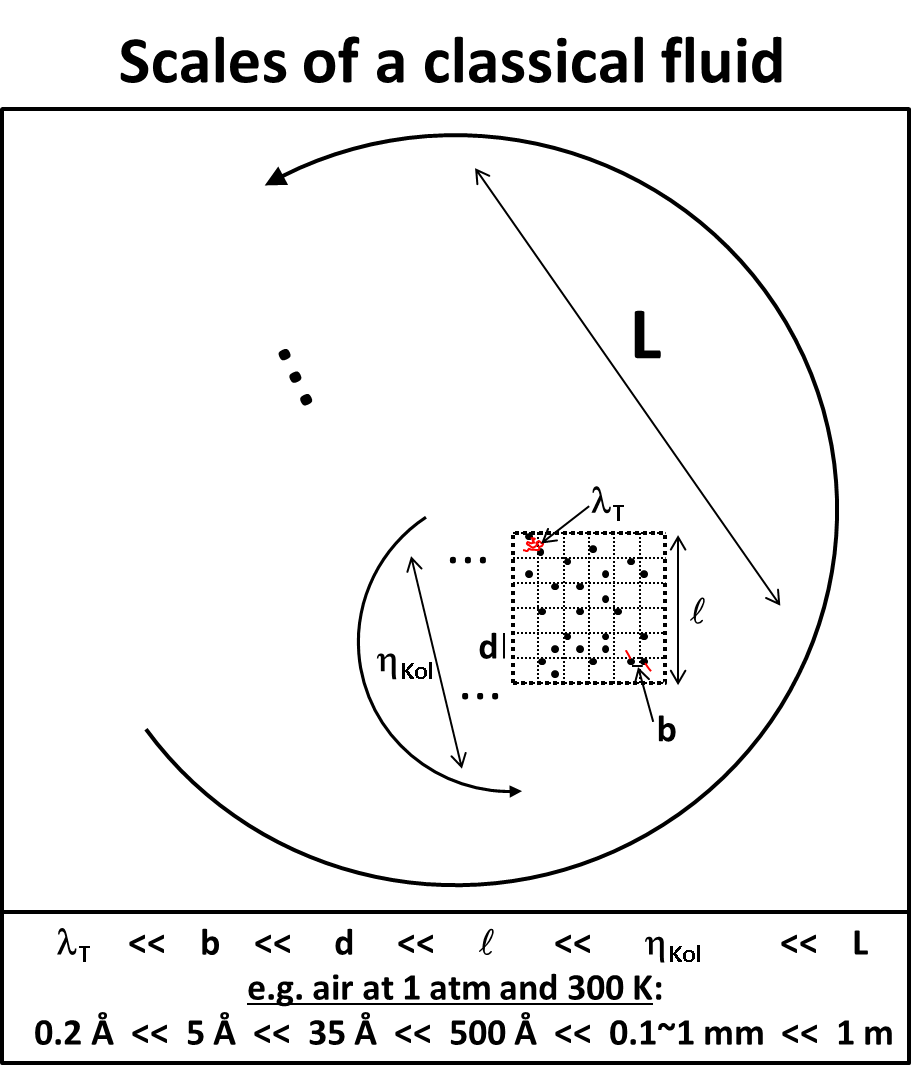} 
\caption{A typical fluid eddy of size $L$ with its macroscopic and microscopic substructure. 
A Kolmogorov-scale sized eddy is the smallest macroscopic structure. 
The successive microscopic substructure is the mean free path $\ell$, the average molecular separation $d$, 
the molecular impact parameter $b$, and the thermal wavelength $\lam_T$. For air at  $1$ atm and room temperature, all of these 
scales are separated by at least an order of magnitude as shown. 
For water, the story is similar but then $b$, $d$, and $\ell$ are all of the same order of magnitude 
of a few Angstroms, 
with the microscopic degrees of freedom strongly coupled. 
Nevertheless, in both cases (air and water) $\lam_T$ is much smaller than 
the average separation between the molecules themselves with their quantum nature therefore inaccessible 
and macroscopically heat appears random. 
In addition, in both cases, the mean free path is much smaller than dissipation scale 
$\eta_{Kol}$, thus validating the continuum approximation, and 
the Navier-Stokes equation becomes the paradigm of interest.}
\label{fig:picture}
\end{figure}

Our primary goal is to study fluid dynamics from a hamiltonian field theory 
point of view evolving its interactions with the similarity renormalization group in order to help elucidate 
the turbulence problem. 
First, however, we start with Newton's equations of motion for a compressible viscous fluid   
in order to better connect with what is already known and to get the correct definition 
of our initial Hamiltonian.

An overview of the paper is now given: 
The conservation equations for a compressible viscous fluid are discussed in a general manner in 
order to set up the rest of the paper. 
The conservation, Euler-Lagrange, and Hamilton sets of equations are shown to be equivalent 
for the ideal fluid and again for the viscous fluid. 
Mass, momentum, and energy conservation 
are shown to give rise to the standard nonholonomic constraint on entropy 
subsequently used to construct the Navier-Stokes Hamiltonian.
To set up the general dynamics of the theory, the final section discusses energy conservation 
of the Navier-Stokes Hamiltonian and derives its dissipative canonical Poisson bracket with  
a general classical dissipative observable. Finally, in two appendices, the classical similarity renormalization group is 
introduced and the symmetries of a nonrelativistic viscous fluid are discussed.

\section{Conservation equations for compressible viscous fluid}
\label{sec:conservation} 

Using Landau and Lifshitz \cite{landau} as a guide, in this section we start with 
mass and momentum conservation, include the principles of local thermodynamics, 
and show that this leads to an energy-entropy equation which upon assuming 
energy conservation produces the standard entropy constraint equation for a 
single-component dissipative fluid. 
This introduces the notation for the rest of the paper, but more importantly, this entropy 
constraint equation is the nonholonomic constraint that is the key to the derivation of $\hns$. 

Mass and momentum conservation of a nonrelativistic fluid are given by \cite{landau}
\bea
\!\!\!\!\!\!\!\!\!\!&\framebox[1in]{mass}&~~~~~~~~~~~~\,\pt\rho+\pdi(\rho v_i)=0\;,\label{eq:1}\\
\!\!\!\!\!\!\!\!\!\!&\framebox[1in]{momentum}&~~~~~\pt(\rho v_i)+\pj(\rho v_iv_j)=\pj\sig_{ij}\label{eq:2}\;,
\eea
where $\rho$ is density, $v_i$ is velocity, and $\sig_{ij}$ is the fluid stress tensor (defined next). 
Repeat indices are summed over the three 
spatial dimensions, $\pt$ is a shorthand for $\frac{\pd}{\pd t}$, and $\pdi$ is a shorthand for $\frac{\pd}{\pd x_i}$. 
$\sig_{ij}$ is the stress tensor of a fluid given by
\be
\sig_{ij}=-p\,\del_{ij}+\sig^\pr_{ij}\;,
\label{eq:3}
\ee
where $p$ is pressure, $\del_{ij}$ is the Kronecker delta, 
and $\sig^\pr_{ij}$ is the viscous stress tensor 
which for a general compressible Navier-Stokes (Newtonian) fluid is defined by 
\bse
\label{eq:sigpr}
\bea
\sig^\pr_{ij}&\equiv&2\eta e_{ij}+\zeta^\pr\Delta\del_{ij}\;,\\
e_{ij}&\equiv&\onehalf (\pdi v_j+\pj v_i)\;,\label{eq:eij}\\
\Delta&\equiv&\pdi v_i=\bn\cdot\bv\;,\label{eq:erate}
\eea
\ese
where $e_{ij}$ is the strain-rate tensor and $\Delta$ is the rate of expansion \cite{batchelor}. 
$\eta$ and $\zeta$ are the shear and bulk viscosity respectively and $\zeta^\pr\equiv\zeta-2\eta/3$, 
defined such that the trace of the stress tensor is independent of shear viscosity.

Multiplying out Eq.~(\ref{eq:2}) and using Eq.~(\ref{eq:1}) leads to the Navier-Stokes equation, 
the dynamical equation for velocity field $\bv(\bx,t)$ with general density $\rho(\bx,t)$:
\be
\fbox{$\pt v_i+v_j\pj v_i=\frac{\pj\sig_{ij}}{\rho}$}\label{eq:2b}\;.
\ee
In summary, momentum conservation is equivalent to both mass conservation 
and the Navier-Stokes equation: pick any two and then you have the other.

Onto energy conservation which implies the entropy constraint of primary interest for the derivation of $\hns$. 
Following Landau and Lifshitz \cite{landau}, first 
take a partial time derivative of the kinetic energy density and then use mass and momentum 
conservation to rearrange the expression. This gives
\beaa
\pt(\onehalf\rho\bv^2)&=&\onehalf(\pt\rho)v_iv_i+\rho v_i(\pt v_i)\\
&=&-\frac{\bv^2}{2}\pj(\rho v_j)+v_i(-\rho v_j\pj v_i+\pj\sig_{ij})\\
&=&-\frac{\bv^2}{2}\pj(\rho v_j)-\rho v_j\pj(\frac{\bv^2}{2})+v_i\pj\sig_{ij}\\
&=&-\pj(\onehalf \rho \bv^2v_j)+v_i\pj \sig_{ij}\;.
\eeaa
Move the first term on the right to the left and we recognize
 the kinetic energy conservation equation of a fluid:
\be
\!\!\!\!\!\!\!\!\!\!\framebox[1.2in]{kinetic energy}~~~~~
\pt(\onehalf\rho\bv^2)+\pdi(\onehalf \rho \bv^2v_i)=v_i\pj \sig_{ij}\label{eq:ke}\;.
\ee
Recall that this followed simply from the product rule of partial differentiation and the mass and 
momentum conservation equations---it is not an independent conservation equation.  
This kinetic energy conservation equation can be described by 
\begin{flushleft}{\it
The net increase in the kinetic energy of a unit volume of fluid per unit time is equal to  
the net flow of kinetic energy into the volume plus the work 
done on the volume by the pressure and viscous forces in the fluid. 
}\end{flushleft}
But there is more, we have not included 
thermal energy yet, the heat part of internal energy. 
Some of the viscous work is converted into heat just like rubbing your hands together 
(but internal to the fluid itself like a self interaction dressing the individual fluid particles). 
This leads to an increase in the entropy of the fluid 
through the second law of thermodynamics as discussed next. 

Internal energy conservation follows from mass conservation and the first and second 
laws of thermodynamics. Strictly speaking the system is not in equilibrium, but locally it is assumed to be. 
Thus as long as we use temporal and spatial differentials, everything still follows the textbook equations of equilibrium thermodynamics. Energy is always conserved, even with dissipation allowed, 
but it changes form as it flows about from work to kinetic energy to heat with entropy of a closed system 
never decreasing. The irreversible part of the 
work increases the entropy according to the second law. This is the physical basis of the relations that follow. 
Using scalar variable $v$ temporarily for specific volume (the inverse of density), 
from the first and second laws of thermodynamics,  locally we have
\bea
du &=& T ds - p dv = T ds +\frac{p}{\rho^2}d\rho\;,\label{eq:td}\\
\Rightarrow \fbox{$\du$}&\equiv&d(\rho u)=\rho du+u d\rho\nn\\
&=&\rho T ds +(u+\frac{p}{\rho})d\rho\nn\\
&=&\fbox{$\rho T ds+h\,d\rho$}\;,\label{eq:rhots}
\eea
where $\cu(\rho,s)\equiv\rho\,u$ is the internal energy density, 
in general a function of specific entropy $s$ and density $\rho$ as this final differential equation shows; 
u is specific internal energy and $h$ 
is specific enthalpy: $h=u+p\,v=u+p/\rho$ 
(the qualifier `specific' implies `per unit total mass' for all quantities in this paper).\footnote{Notation 
warning: `$h$' in fluid dynamics is often helicity or helicity density, however 
in this paper $h = u + p/\rho$ is the specific enthalpy.}
The above relations for $\du$ and $du$ imply the 
following definitions for temperature and pressure:
\bea
T=\left(\frac{\pd u}{\pd s}\right)_{\rho}&=&
\frac{1}{\rho}\left(\frac{\pd\cu}{\pd s}\right)_{\rho}\;,\label{eq:tdef}\\
p=\rho^2\left(\frac{\pd u}{\pd \rho}\right)_{s}&=&
\rho\left(\frac{\pd\cu}{\pd \rho}\right)_{s}-\cu\;.\label{eq:pdef}
\eea
So, $\cu(\rho,s)$, the internal energy density, is where the temperature and pressure effects 
implicitly lie in the Hamiltonians that follow.  
Before leaving this part of the derivation, 
note the following standard relation that will be used later as well:
\be
dh = d(u+p v)=T ds + v dp = T ds + dp/\rho\;.
\label{eq:dh}
\ee
For any instant in time, locally in space, Eq.~(\ref{eq:dh}) implies that 
the ``$\bn p/\rho$'' term of fluid dynamics can be replaced by
\be
\frac{\bn p}{\rho}=\bn h-T\bn s\;.
\label{eq:gradh}
\ee

Continuing like above with the kinetic energy, but now for the internal energy:  
take a partial time derivative of the internal energy density and use mass 
conservation as well as the thermodynamic relations just discussed to rearrange the expression 
leaving the internal energy conservation equation. Thus we have 
\beaa
\pt\cu&\stackrel{(\ref{eq:rhots})}{=}&\rho T \pt s+h \pt\rho\\
&\stackrel{(\ref{eq:1})}{=}&\rho T \pt s-h \pdi(\rho v_i)\\
&=&\rho T \pt s+\rho v_i \pdi h-\pdi(\rho h v_i)\\
&\stackrel{(h=u+p/\rho)}{=}&\rho T \pt s+\rho v_i \pdi h-\pdi(\cu v_i)-\pdi(p v_i)\;.
\eeaa
Moving the internal energy flux term to the left and rearranging gives
\beaa
\pt\cu+\pdi(\cu v_i)&=&\rho T \pt s+\rho v_i \pdi h-\pdi(p v_i)\\
&\stackrel{(\ref{eq:gradh})}{=}&\rho T \pt s+\rho T v_i \pdi s+v_i \pdi p-\pdi(p v_i)\\
&=&\rho T D_t s-p \Delta\;,
\eeaa
where
\be
\fbox{$D_t\equiv\pt+\bv\cdot\bn$}\label{eq:dt}
\ee
is the standard material derivative  
and recall $\Delta$ is Batchelor's 
shorthand for the rate of expansion, Eq.~(\ref{eq:erate}). 
In summary, internal energy conservation is given by
\be
\!\!\!\!\!\!\!\!\!\!\framebox[1.2in]{internal energy}~~~~~
\pt\cu+\pdi(\cu v_i)=\rho T D_t s-p \Delta\label{eq:u}\;.
\ee
This followed from mass conservation and the local laws of thermodynamics. 
It is an independent dynamical equation for the internal energy density and is an expression 
of the first and second laws of thermodynamics applied to a fluid as it flows about. 
To help with interpretation of the quantities, note that Eq.~(\ref{eq:u}) can be exactly rearranged  
by expanding out its second term on the left (the flux term). This gives
\bee
D_t\cu=\rho T D_t s-\rho h \Delta\;,
\eee
where $\rho\,h=p+\cu$ is the enthalpy density. Thus we see this really is just 
``$du=T ds-p\,dv$'' (with mass conservation included) along the pathlines of a fluid. 
Enthalpy density acts like the pressure ``p'' here and 
$\Delta$ acts like the ``$dv$'' which is as it should be. On page 75 
of Batchelor \cite{batchelor} from the divergence theorem, 
$\Delta$ is the rate of relative volume change: 
\bee
\Delta=\bn\cdot\bv=\frac{1}{d^3x}\sum\bv\cdot \hat{n}\,dA=
\lim_{\del Vol\rightarrow 0}\sum\frac{1}{\del Vol}\frac{\del Vol_{\hat{n}}}{\del t}\;,
\eee
or the ``local rate of expansion'' as Batchelor calls it.

Now we have the two required pieces for the energy-entropy equation. 
Putting it all together: add kinetic energy conservation, Eq.~(\ref{eq:ke}), 
with internal energy conservation, Eq.~(\ref{eq:u}). 
The left-hand sides add simply as written below in Eq.~(\ref{eq:energyentropy}). The right-hand 
sides are exactly rearranged as 
\bea
RHS_{KE+\cu}&=&
v_i\pj \sig_{ij}+\rho T D_t s-p \Delta\nn\\
&\stackrel{(\ref{eq:3})}{=}&-v_i\pdi p+v_i\pj \sig_{ij}^\pr+\rho T D_t s-p\pdi v_i\nn\\
&\stackrel{({\rm same}~p)}{=}&-\pdi(v_i p)+v_i\pj \sig_{ij}^\pr+\rho T D_t s\nn\\
&=&-\pdi(v_i p)+\pj(v_i \sig_{ij}^\pr)-\sig_{ij}^\pr\pj v_i+\rho T D_t s\nn\\
&\stackrel{(\ref{eq:3})}{=}&\pj(v_i \sig_{ij})-\sig_{ij}^\pr\pj v_i+\rho T D_t s\;.\label{eq:rhs}
\eea
Note carefully that $\sig_{ij}^\pr$ is the viscous stress tensor, whereas $\sig_{ij}$ is the 
full stress tensor which 
includes pressure too according to Eq.~(\ref{eq:3}).  
The point of obtaining this last equation, Eq.~(\ref{eq:rhs}), is that this first term 
on the right can be physically identified with an energy flux term as discussed next. 

Thus, altogether combining kinetic energy and internal energy conservation, 
and noting that the right-hand side can be rearranged as in Eq.~(\ref{eq:rhs}), 
we are left with the energy-entropy equation:
\be
\fbox{$
\pt(\cu+\onehalf\rho\bv^2)+\pdi(\cu v_i+\onehalf \rho \bv^2v_i)=
\pj(v_i \sig_{ij})+\left[\rho T D_t s-\sig_{ij}^\pr\pj v_i\right]$}\;.
\label{eq:energyentropy}
\ee
The square brackets group the entropy related terms and are convenient for later discussions. 
Note that Eq.~(\ref{eq:energyentropy}) is the same as in Landau and Lifshitz near the bottom of
p.~193 \cite{landau}, except that the heat flux term has not been added and subtracted yet---we do this below. 
Also, Landau and Lifshitz did not provide separate kinetic and internal 
energy relations as we have done above with Eqs.~(\ref{eq:ke}) and (\ref{eq:u}). 
We find these separate kinetic and internal energy relations instructive and thus 
have included them here. 
Recall how it all came together above: this energy-entropy equation came from mass and 
momentum conservation and the first and second laws of local thermodynamics. 
It is one further dynamical equation beyond mass and momentum conservation 
because thermodynamics (thermal energy) has now been consistently taken into account. 

As hinted by its joint name, the energy-entropy equation contains two physical principles within it: 
energy conservation (temporal translational invariance) 
and the increase of entropy due to irreversible work (second law of thermodynamics). 
The energy conservation terms in Eq.~(\ref{eq:energyentropy}) 
must be supplemented by a heat flux term that accounts for thermal conduction in the absence of motion but 
with inhomogeneities in the fluid due to temperature or chemical potential gradients. Thus, 
following Landau and Lifshitz \cite{landau}, a ``$\bn\cdot\bq$'' term is added and subtracted to the energy-entropy equation. 
$\bq$ is the heat flux which is the heat energy per unit time per unit area from heat conduction in the fluid. In the case without 
diffusion (our initial interest) $\bq=-\kappa\,\bn T$ where $\kappa$ is the thermal conductivity. 
Given this, as motivated from Eq.~(\ref{eq:energyentropy}) itself, 
the energy conservation equation for a nonrelativistic viscous fluid is defined by \cite{landau}
\be
\!\!\!\!\!\!\!\!\!\!\framebox[1in]{energy}~~~~~
\pt(\cu+\rho\bv^2/2)+\pdi(v_i\,\cu+v_i\,\rho \bv^2/2)\equiv
\pj\left(v_i \sig_{ij}\right)-\bn\cdot\bq\;,
\label{eq:energy}
\ee
which in words can be stated as 
\begin{flushleft}{\it
The net increase in the kinetic-plus-internal energy of a unit volume of fluid per unit time is equal to 
the net flow of kinetic-plus-internal energy into the volume plus the heat conducted into the same volume  
and the net work done by the pressure and viscous forces on its surface.
}\end{flushleft}
Bringing everything consistently to the left-hand side, as described on p.\ 229 of Landau and Lifshitz \cite{landau},  
$q_i-v_j\,\sig_{ji}^\pr$ is the irreversible energy flux of a fluid 
whereas the remaining terms in the divergence part of Eq.~(\ref{eq:energy})---including 
the pressure term---make up the reversible energy flux: 
$v_i\,\cu+v_i\rho \bv^2/2+v_i\,p=v_i\rho h+v_i\rho \bv^2/2$. 
Interestingly, note that the energy density term of Eq.~(\ref{eq:energy}), $\cu+\onehalf\rho\bv^2$, 
ends up being the final form of $\che$ {\it and} $\chns$, the very Hamiltonians we are seeking. 
But the point is that $\bv(\bx,t)$ has not been defined yet in terms of Hamiltonian variables with a 
dissipative canonical Poisson bracket structure \cite{hiroki,zakharov}. This 
will be done in the respective sections that follow. 

Given energy conservation, Eq.~(\ref{eq:energy}), then the energy-entropy equation, Eq.~(\ref{eq:energyentropy}), becomes 
the entropy terms in square brackets along with the $\bn\cdot\bq$ heat flux term with the correct sign. 
As was to be derived, we are left with the {\it entropy constraint equation}:
\be
\fbox{$\rho T D_t s=\sig_{ij}^\pr\pj v_i-\bn\cdot\bq$}\;.\label{eq:scon}
\ee
Summarizing the story that led up to this entropy constraint equation: it came simply from  
mass, momentum, and energy conservation applied to a classical nonrelativistic viscous fluid 
where the concept of energy was necessarily enlarged to include heat 
according to the standard laws of local equilibrium thermodynamics. 
In the words of Landau and Lifshitz \cite{landau}: the left-hand side of Eq.~(\ref{eq:scon}) is the 
``amount of heat gained by unit volume of the fluid,'' the first term on the right is the 
``energy dissipated into heat by viscosity,'' and the last term 
is the ``heat conducted into the volume concerned.''

Before leaving this section we have two further notes: 
First, as can be seen by perusing the explicit first-principle proofs above, 
nowhere in these derivations was the actual form of the viscous stress tensor, $\sig_{ij}^\pr$, 
used.\footnote{e.g.\ we did not assume that $\sig_{ij}^\pr$ was symmetric---although with angular momentum conservation 
it is \cite{hiroki}---or more importantly we did not assume 
that its form was necessarily that of a Newtonian fluid.}
Thus, {\it the above entropy constraint equation also holds for a non-Newtonian fluid}. However, 
we will still call our results of the final section the ``Navier-Stokes Hamiltonian'' (defined by the 
Newtonian viscous stress tensor of Eq.~(\ref{eq:sigpr})) because we think this  
is already interesting in itself, but keep in mind that 
the final $H_{NS}$ is actually more generally applicable.
Second, as seen in the middle line with the ``same p'' comment above the equal sign 
in the block of equations ending with Eq.~(\ref{eq:rhs}),
the following two pressures where assumed to be equivalent: the thermodynamic one from
\bee
du = T ds - p \,dv
\eee
(the so-called equilibrium pressure \cite{batchelor}), and the one from the Navier-Stokes equation,  
\bee
\sig_{ij}\equiv-p\,\del_{ij}+\sig_{ij}^\pr\;.
\eee
In other words we are assuming that the pressure of the ``$\bn p/\rho$'' term in the Navier-Stokes equation 
is equivalent to the local equilibrium pressure of thermodynamics. For clarity note however that we are 
including a nonzero bulk viscosity $\zeta$. Therefore our pressure $p$ is not the same as the normal pressure 
acting on an arbitrary surface. This follows by noting (recall the definitions of Eq.~(\ref{eq:sigpr}))
\bea
p_{normal}&\equiv&-\frac{1}{3}{\rm tr}(\sig)=-\frac{1}{3}(-3p+2\eta\Delta+3\zeta^\pr\Delta)\nn\\
&=&-\frac{1}{3}\left[-3p+2\eta\Delta+3\zeta\Delta-3\left(\frac{2}{3}\eta\right)\Delta\right]\nn\\
&=&p-\zeta\Delta\;.\label{eq:pnormal}
\eea
So the pressure in the Navier-Stokes equation and the pressure normal to an 
arbitrary surface in the fluid differ by a term first order in $\zeta\Delta$; 
this includes a velocity derivative and a factor of the bulk viscosity. In 
nearly-incompressible fluids such as water it may be hard to discern a difference between 
$p$ and $p_{normal}$ and so we will not worry about this distinction for now, 
but note $\zeta$ and $\Delta$ remain arbitrary fields at this point and therefore the physics 
of Eq.~(\ref{eq:pnormal}) is contained in what follows.

\section{Euler Hamiltonian}
\label{sec:euler}

As a warm up for deriving $\hns$ without the complication of dissipation, but also because 
the ``no viscosity'' and ``small viscosity'' theories really are different beasts, 
in this section we derive the Hamiltonian corresponding 
to the Euler equation of a general ideal fluid.  
By `general' we mean the internal energy density depends on both density and specific entropy, 
and by `ideal' we mean the usual semantics of the viscosity terms of the Navier-Stokes equation being 
dropped from the beginning. 

The procedure for deriving the Euler Hamiltonian $H_E$ starts with a definition of the canonical lagrangian density for 
a general ideal fluid according to Zakharov and Kuznetsov \cite{zakharov}:
\be
\fbox{$
\cle=\onehalf \rho\bv^2-\cu(\rho,s)+\phi\,[\pt\rho+\pdi(\rho v_i)]+\alpha\,(\pt\beta+v_i\pdi\beta)
+\lam\,(\pt s+v_i\pdi s)$}\;.\label{eq:leuler}
\ee
The variables of $\cle$ are described next, but first note that compared to \cite{zakharov} 
we change the names of some of the variables as well as 
the sign of the $\alpha$ and $\lam$ lagrange multiplier scalar fields 
so that they are related by a plus sign to the conjugate momentum field that they represent 
($\pi_\beta=+\alpha$ and $\pi_s=+\lam$). 
The variables of $\cle$ are
\beaa
\rho&:&{\rm density}\\
\bv&:&{\rm constrained}~{\rm velocity}~{\rm field}\\
\cu&:&{\rm internal}~{\rm energy}~{\rm density}\\
s&:&{\rm specific}~{\rm entropy}\\
\phi&:&{\rm velocity}~{\rm potential}\\
(\alpha,\beta)&:&{\rm Clebsch}~{\rm potential}~{\rm pair}\\
\lam&:&{\rm lagrange}~{\rm multiplier}~{\rm field}~{\rm for}~{\rm the}~{\rm ideal}~{\rm entropy}~{\rm constraint}
\eeaa

Now we derive the Euler-Lagrange equations for $\cle$ and show that they are 
equivalent to the equations of motion of an ideal fluid: the continuity, Euler,   
and ideal entropy constraint equations. Then the section concludes with a derivation of $H_E$ and shows that 
its Hamilton equations satisfy these same dynamics.

\subsection{Euler-Lagrange equations for ideal fluid}

The Euler-Lagrange equation (for derivation, see Section~\ref{sec:ELeq}) for arbitrary scalar field $\Phi$, from 
a lagrangian density with at most first order partial derivatives (the usual case), is given by \cite{HT}
\be
\pt\left(\frac{\pd\cl}{\pd(\pt\Phi)}\right)+\pdi\left(\frac{\pd\cl}{\pd(\pdi\Phi)}\right)=\frac{\pd\cl}{\pd\Phi}
\;.
\ee
Thus the Euler-Lagrange equations from varying all the independent 
fields of $\cle$ in Eq.~(\ref{eq:leuler}) are as follows.

\underline{Varying $\rho$}:
\bea
\pt\left(\frac{\pd\cle}{\pd(\pt\rho)}\right)&+&\pdi\left(\frac{\pd\cle}{\pd(\pdi\rho)}\right)=\frac{\pd\cle}{\pd\rho}\;,\nn\\
\Rightarrow~~\pt\phi&+&\pdi\left(\phi v_i\right)~~~~~~\,=\frac{\bv^2}{2}-\frac{\pd\cu}{\pd\rho}+\phi\pdi v_i\nn\\
&&~~~~~~~~~~~~~~~\,\stackrel{(\ref{eq:pdef})}{=}\frac{\bv^2}{2}-h+\phi\pdi v_i\;,\nn\\
\Rightarrow&&\fbox{$D_t\phi-\frac{\bv^2}{2}+h=0$}\;,\label{eq:euler1}
\eea
where $D_t$ is the material derivative of Eq.~(\ref{eq:dt}). 
This is the correct dynamical equation for velocity potential $\phi$ as shown below. 
The velocity field itself, $\bv$ from $\cle$, satisfies a constraint equation with no partial time derivative as follows 
from its Euler-Lagrange equation:

\underline{Varying $\bv$}:
\bea
\pt\left(\frac{\pd\cle}{\pd(\pt v_i)}\right)&+&\pj\left(\frac{\pd\cle}{\pd(\pj v_i)}
\right)=~~\frac{\pd\cle}{\pd v_i}\;,\nn\\
\Rightarrow~~\pt\left(0\right)&+&\pj\left(\rho\phi\del_{ij}\right)~~~~~=~~\rho v_i
+\phi\pdi\rho+\alpha\pdi\beta+\lam\pdi s\;,\nn\\
\Rightarrow &&\fbox{$\bv=\bn\phi-\frac{\alpha}{\rho}\bn\beta-\frac{\lam}{\rho}\bn s$}\;.\label{eq:v}
\eea
Often in what follows we keep writing ``$\bv$'' for simplicity, but 
what is really meant is the right-hand side of this constraint 
equation, which has the dynamical fields of interest. 
This includes the dynamical equation for $\phi$ that was just derived, Eq.~(\ref{eq:euler1}), 
including of course the $\bv$ inside the material derivative. 
Now varying the remaining fields of $\cle$ gives

\underline{Varying $\phi$}:
\be
\Rightarrow~~~~\fbox{$\pt\rho+\bn\cdot(\rho\bv)=0$}\;;\label{eq:24}
\ee

\underline{Varying $\alpha$}:
\be
\Rightarrow~~~~\fbox{$D_t\beta=0$}\;;
\ee

\underline{Varying $\beta$}:
\bea
\Rightarrow\pt\alpha&+&\pdi\left(\alpha v_i\right)~=~0\;,\nn\\
\Rightarrow &&\fbox{$\pt\alpha+\bn\cdot(\alpha\bv)=0$}\;;
\eea

\underline{Varying $\lam$}:
\be
\Rightarrow~~~~\fbox{$D_t s=0$}\;;
\ee

\underline{Varying $s$}:
\bea
\Rightarrow\pt\lam&+&\pdi\left(\lam v_i\right)~=~-\frac{\pd\cu}{\pd s}\nn\\
&&~~~~~~~~~~\,\stackrel{(\ref{eq:tdef})}{=}~-\rho T\;,\nn\\
\Rightarrow &&\fbox{$\pt\lam+\bn\cdot(\lam\bv)+\rho T=0$}\;.\label{eq:euler7}
\eea

In summary, varying $\cle$ of Eq.~(\ref{eq:leuler}) produces one constraint equation for vector field $\bv$ 
and six equations of motion for scalar fields $\rho$, $\phi$, $\alpha$, $\beta$, $\lam$, and $s$:  
Eqs.~(\ref{eq:euler1})--(\ref{eq:euler7}).  
Now we show that these scalar equations of motion are equivalent to 
the dynamics contained in the Euler equation. Then we move 
on to the Euler Hamiltonian and show how its  
resulting Hamilton equations reproduce these same equations of motion.

\subsection{Euler equation equivalence proof}

The exact equivalence between the Euler-Lagrange equations of $\cle$ and 
the Euler equation (with mass conservation and basic thermodynamics included as well) 
is shown here. 
Starting with the momentum conservation equation of an ideal fluid, we insert velocity constraint 
Eq.~(\ref{eq:v}), use thermodynamic relation Eq.~(\ref{eq:gradh}), 
and show that the result can be exactly rearranged into a form proportional to the six dynamical equations of 
motion just discussed for $\rho$, $\phi$, $\alpha$, $\beta$, $\lam$, and $s$. 

\subsubsection{Mass and momentum conservation equivalence proof for ideal fluid}

Start with momentum conservation operator Eq.~(\ref{eq:2}) with all terms moved to the left and 
the viscosity terms dropped:
\be
\coe\equiv\pt(\rho v_i)+\pj(\rho v_i v_j)+\pdi p\;;\label{eq:oe}
\ee
substitute velocity constraint Eq.~(\ref{eq:v}) written as
\be
\rho v_i=\rho\pdi\phi-\alpha\pdi\beta-\lam\pdi s\;;
\ee
use thermodynamic relation Eq.~(\ref{eq:gradh}):
\be
\pdi p=\rho\pdi h-\rho T \pdi s\;;
\ee
and finally exactly rearrange the result into a form proportional to the dynamical Euler-Lagrange equations. 
This follows as shown below.  
Note that we are not setting $\coe$ to zero at this point, but rather are seeing how its operator form varies 
with these six scalar potentials. 
After this $\bv$ constraint and thermodynamic $\bn p$ substitution has been made, the above $\coe$ becomes
\be
\coe\rightarrow\pt(\rho\pdi\phi-\alpha\pdi\beta-\lam\pdi s)+
\pj\left[v_j(\rho\pdi\phi-\alpha\pdi\beta-\lam\pdi s)\right]+
\rho\pdi h-\rho T \pdi s\;.
\ee
This looks a little unwieldy, but we see with the product rule of partial differentiation, 
this will only produce 17 terms. Continuing: expand out the differential products, collect common factors, 
and exactly rearrange by adding and subtracting identical terms to get the $D_t\beta$ and $D_t s$ terms to work out 
right; then note three cancellations and the remaining dust becomes the $\bv^2/2$ 
term because of the $\bv$ constraint consistently upheld. Thus we obtain 
\bea
\coe~=&&\rho\pdi h-\rho T \pdi s+\nn\\
&+&(\pt\rho)\pdi\phi+\rho\pdi\pt\phi-(\pt\alpha)\pdi\beta-\alpha\pdi\pt\beta-(\pt\lam)\pdi s-\lam\pdi\pt s\nn\\
&+&(\pj v_j)(\rho\pdi\phi-\alpha\pdi\beta-\lam\pdi s)+v_j(\pj\rho)\pdi\phi+v_j\rho\pdi\pj\phi\nn\\
&-&v_j(\pj\alpha)\pdi\beta-v_j\alpha\pdi\pj\beta-v_j(\pj\lam)\pdi s-v_j\lam\pdi\pj s\nn\\
=&&(\pdi\phi)\left[\pt\rho+\bn\cdot(\rho\bv)\right]\nn\\
&-&(\pdi\beta)\left[\pt\alpha+\bn\cdot(\alpha\bv)\right]\nn\\
&-&(\pdi s)\left[\pt\lam+\bn\cdot(\lam\bv)+\rho T\right]\nn\\
&-&\alpha\pdi\left[D_t\beta\right]\nn\\
&-&\lam\pdi\left[D_t s\right]\nn\\
&+&\rho\pdi\left[D_t\phi-\frac{\bv^2}{2}+h\right]\;.\label{eq:magic}
\eea
Seemingly almost like magic (symmetry being the underlying magician), 
we see $\coe$ is proportional to all six of these dynamical Euler-Lagrange equations under discussion. 
Finally note the following simple result:
\bea
\coe&\equiv&\pt(\rho v_i)+\pj(\rho v_i v_j)+\pdi p\nn\\
&=&v_i\left[\pt\rho+\bn\cdot(\rho\bv)\right]+\rho\left[\pt v_i+v_j\pj v_i+\frac{\pdi p}{\rho}\right]\;,
\eea
which as mentioned earlier, shows that the momentum conservation operator is proportional 
to the continuity and in this case Euler equation operator. 
Thus, if all six of these independent Euler-Lagrange equations are satisfied (with respective operator vanishing), 
then $\coe=0$ and the Euler and continuity equations must therefore also both be satisfied and the two 
approaches are therefore dynamically equivalent. q.e.d. A final remark is to recall that in order to 
obtain this exact equivalence, local equilibrium thermodynamics was used too through Eq.~(\ref{eq:gradh}). 
In short, $\cle$ of Eq.~(\ref{eq:leuler}) has the same dynamics as the general Euler equation where 
pressure depends on density and entropy. Now we use $\cle$ to derive the Euler Hamiltonian.

\subsection{Hamilton equations for ideal fluid}

This section completes the derivation of the Euler Hamiltonian, $H_E$, and shows 
that its Hamilton equations are the same six dynamical equations of motion as 
from the Euler-Lagrange equations just discussed.  
These results are well known, but the final section with the Navier-Stokes Hamiltonian is  
not as well known and the two procedures are compared in the end to highlight 
the differences and similarities between obtaining a Hamiltonian with and without dissipation. 

To derive $H_E$, first we need to determine its nonzero conjugate momentum fields. 
These follow simply from $\cle$ of Eq.~(\ref{eq:leuler}): 
\bea
\pi_\rho&\equiv&\frac{\pd\cle}{\pd(\pt\rho)}=\phi(\bx,t)\;,\\
\pi_\beta&\equiv&\frac{\pd\cle}{\pd(\pt\beta)}=\alpha(\bx,t)\;,\\
\pi_s&\equiv&\frac{\pd\cle}{\pd(\pt s)}=\lam(\bx,t)\;.
\eea
In the following, since it is unambiguous and leaves a cleaner notation, often we will continue to 
write `$\phi$', `$\alpha$', and `$\lam$' for the conjugate momentum fields, although this is actually 
a shorthand for the conjugate momentum fields they represent: 
`$\pi_\rho(\bx,t)$', `$\pi_\beta(\bx,t)$', and `$\pi_s(\bx,t)$' 
respectively.

Given these conjugate momenta, $H_E$ follows from the canonical Hamiltonian 
procedure\footnote{The canonical Hamiltonian procedure with dissipation 
is derived in Section~\ref{sec:hameq} of this paper.} \cite{HT}:
\bea
H_E&\equiv&\intx\che\;,\label{eq:heint}\\
\che&=&\pi_\rho\pt\rho+\pi_\beta\pt\beta+\pi_s\pt s-\cle\nn\\
&=&\phi\pt\rho+\alpha\pt\beta+\lam\pt s-\cle\nn\\
&=&-\onehalf\rho\bv^2+\cu(\rho,s)-\phi\,\bn\cdot(\rho\bv)-\alpha\,(\bv\cdot\bn)\beta-\lam\,(\bv\cdot\bn)s
\;.\label{eq:h1}
\eea
Note $\bv$ satisfies the same constraint that came from varying $\cle$ 
which for emphasis is rewritten here. `$\bv$' 
in Eq.~(\ref{eq:h1}) is really a shorthand for
\be
\fbox{$
\bv\rightarrow\bn\phi-\frac{\alpha}{\rho}\bn\beta-\frac{\lam}{\rho}\bn s$}\;.\label{eq:v2}
\ee
Thus, these right three terms of Eq.~(\ref{eq:h1}) are seen to be related to the $\bv^2$ term. 
With a slight rearrangement of the third to last term, the above becomes 
\bea
\che&=&-\onehalf\rho\bv^2+\cu(\rho,s)-\pdi(\rho \phi v_i)+\rho v_i\pdi\phi
-\alpha v_i\pdi\beta-\lam v_i\pdi s\nn\\
&=&-\onehalf\rho\bv^2+\cu(\rho,s)+\rho v_i\left(\pdi\phi-\frac{\alpha}{\rho}\pdi\beta-\frac{\lam}{\rho}\pdi s\right)
-\pdi(\rho \phi v_i)\nn\\
&\stackrel{(\ref{eq:v2})}{=}&-\onehalf\rho\bv^2+\cu(\rho,s)+\rho \bv^2-\bn\cdot(\rho \phi \bv)\nn\\
&=&\onehalf\rho\bv^2+\cu(\rho,s)-\bn\cdot(\rho \phi \bv)\;.\label{eq:h2}
\eea
Upon performing the integration required by Eq.~(\ref{eq:heint}), the last term of Eq.~(\ref{eq:h2}) 
is seen to be a surface integral at spatial infinity which vanishes   
on the physical grounds of all finite-energy fields vanishing there. Thus the Euler Hamiltonian 
for a general ideal fluid is given by
\be
\fbox{$
H_E[\rho, \phi; \beta, \alpha; s, \lam]=\intx\left[\onehalf\rho\bv^2
+\cu(\rho,s)\right]$}_{~{\rm with}~{\rm constraint}~(\ref{eq:v2})}\;.\label{eq:he}
\ee
Note that these functional arguments of $H_E$ on the left are 
written in terms of its three dynamical coordinate fields: $\rho$, $\beta$, and $s$, and 
their respective conjugate momentum fields: $ \phi$, $\alpha$, and $\lam$.

Now onto Hamilton's equations. Out of clarity, in the Navier-Stokes Hamiltonian section that follows, we carefully 
go through the derivation of Hamilton's equations because they are modified due to dissipation. 
However for this ideal fluid let us just quote the standard field theory result 
for Hamilton's equations of a Hamiltonian with coordinate scalar field $\Phi$, for example, and  
conjugate momentum field $\pi_\Phi$ \cite{HT}:
\bse
\label{eq:ehe}
\bea
\pt\Phi&=&\frac{\del\che}{\del\pi_\Phi}=\frac{\pd\che}{\pd\pi_\Phi}
-\pdi\left(\frac{\pd\che}{\pd(\pdi\pi_\Phi)}\right)\;,\\
\pt\pi_\Phi&=&-\frac{\del\che}{\del\Phi}=-\frac{\pd\che}{\pd\Phi}
+\pdi\left(\frac{\pd\che}{\pd(\pdi\Phi)}\right)\;.
\eea
\ese
The opposite sign of the last terms on the right follows from a spatial
integration by parts and the dropping of spatial boundary terms at infinity because 
all field variations $\del\Phi$ are assumed to vanish there. 
Note also that these Hamilton equations are based on variations of the field 
$\del\Phi$ vanishing at the temporal endpoints (required for 
the Euler-Lagrange equation derivation itself which is used under-the-hood 
in defining a Hamiltonian \cite{landau2}). 
We mention these boundary condition details to emphasize the fact 
that they cause no conceptual problem for the ideal fluid, but with dissipation necessitate 
introducing initial-position vector field $\ba(\bx,t)$ as the dynamical coordinate field 
of the Navier-Stokes Hamiltonian. 
Thus, this is where the differences between the Euler 
and Navier-Stokes Hamiltonians originate. 
  
Thus, the Hamilton equations for 
the three respective coordinate pairs of $H_E$ are 

\underline{coordinate field $\rho$}:
\bea
\pt\rho&=&\frac{\del\che}{\del\pi_\rho}=\frac{\del\che}{\del\phi}=\frac{\pd\che}{\pd\phi}
-\pdi\left(\frac{\pd\che}{\pd(\pdi\phi)}\right)\;,\nn\\
\Rightarrow\pt\rho&=&-\pdi(\rho v_i)\;,\nn\\
\Rightarrow&&\fbox{$\pt\rho+\bn\cdot(\rho\bv)=0$}\;;
\eea

\underline{conjugate momentum field $\pi_\rho=\phi$}:
\bea
\pt\pi_\rho&=&\pt\phi=-\frac{\del\che}{\del\rho}=-\frac{\pd\che}{\pd\rho} 
+\pdi\left(\frac{\pd\che}{\pd(\pdi\rho)}\right)\;,\nn\\
\Rightarrow\pt\phi&=&~-\onehalf\bv^2
-\frac{\rho}{2}2v_i\left[\frac{1}{\rho^2}\left(\alpha\pdi\beta+\lam\pdi s\right)\right]
-\frac{\pd\cu}{\pd\rho}\nn\\
&\stackrel{(\ref{eq:v2})}{=}&~+\onehalf\bv^2-\bv\cdot\bn\phi
-\frac{\pd\cu}{\pd\rho}\nn\\
&\stackrel{(\ref{eq:pdef})}{=}&~~\onehalf\bv^2-\bv\cdot\bn\phi-h\;,\nn\\
\Rightarrow&&\fbox{$D_t\phi-\frac{\bv^2}{2}+h=0$}\;;
\eea

\underline{coordinate field $\beta$}:
\bea
\pt\beta&=&\frac{\del\che}{\del\pi_\beta}=\frac{\del\che}{\del\alpha}=\frac{\pd\che}{\pd\alpha}
-\pdi\left(\frac{\pd\che}{\pd(\pdi\alpha)}\right)\;,\nn\\
\Rightarrow\pt\beta&=&v_i(-\pdi\beta)\;,\nn\\
\Rightarrow&&\fbox{$D_t\beta=0$}\;;
\eea

\underline{conjugate momentum field $\pi_\beta=\alpha$}:
\bea
\pt\pi_\beta&=&\pt\alpha=-\frac{\del\che}{\del\beta}=-\frac{\pd\che}{\pd\beta}
+\pdi\left(\frac{\pd\che}{\pd(\pdi\beta)}\right)\;,\nn\\
\Rightarrow\pt\alpha&=&\pdi[v_i(-\alpha)]\;,\nn\\
\Rightarrow&&\fbox{$\pt\alpha+\bn\cdot(\alpha\bv)=0$}\;;
\eea

\underline{coordinate field $s$}:
\bea
\pt s&=&\frac{\del\che}{\del\pi_s}=\frac{\del\che}{\del\lam}=\frac{\pd\che}{\pd\lam}
-\pdi\left(\frac{\pd\che}{\pd(\pdi\lam)}\right)\;,\nn\\
\Rightarrow\pt s&=&v_i(-\pdi s)\;,\nn\\
\Rightarrow&&\fbox{$D_t s=0$}\;;
\eea

\underline{conjugate momentum field $\pi_s=\lam$}:
\bea
\pt\pi_s&=&\pt\lam=-\frac{\del\che}{\del s}=-\frac{\pd\che}{\pd s}
+\pdi\left(\frac{\pd\che}{\pd(\pdi s)}\right)\;,\nn\\
\Rightarrow\pt\lam&=&-\frac{\pd\cu}{\pd s}+\pdi[v_i(-\lam)]\nn\\
&\stackrel{(\ref{eq:tdef})}{=}&~~-\rho T-\bn\cdot(\lam\bv)\;,\nn\\
\Rightarrow&&\fbox{$\pt\lam+\bn\cdot(\lam\bv)+\rho T=0$}\;.
\eea
These are the same as the above Euler-Lagrange equations for an ideal fluid. q.e.d.\

\section{Navier-Stokes Hamiltonian}
\label{sec:ns} 

This section in a sense is simply a repeat of the previous section with viscosity included. 
However, this one simple fact leads to dissipation and complicates the derivation of a Hamiltonian. 
Nevertheless, it is straightforward as shown by this paper. This section derives 
the canonical Navier-Stokes Hamiltonian, $\hns$, from first principles. 
In order to include dissipation, entropy constraint Eq.~(\ref{eq:scon}) must be accounted for consistently. 
As emphasized by Fukagawa and Fujitani \cite{hiroki}, 
the viscosity dependence of $\hns$ must enter through variations of the specific entropy field, 
$\del s(\bx,t)$, and this necessarily introduces a nonholonomic (path-dependent) constraint \cite{goldstein} on the system. 
In order to treat this constraint properly, with variations in entropy 
being proportional to variations in a coordinate (field), one must 
introduce initial-position vector field $\bax$ \cite{hiroki}, which 
is a label to all of the fluid particles \cite{zakharov} and  
physically given by their initial positions at some arbitrary time $t_0$. 
Another way to say it is that $\bax$ is the inverse 
mapping of lagrangian coordinate $\bx(\ba,t)$. 
We are still in the Euler representation of a fluid with 
the standard local field theory point of view being maintained and  
$\bax$ denotes the initial position of  
the fluid particle that happens to be at $\bx$ at time $t$. 
As shown below, $\bax$ turns out to be the main dynamical coordinate field of $\hns$ 
and is assumed to be a one-to-one map to all of their initial positions \cite{zakharov}. 
Although some of phase space could be missed by this one-to-one assumption, in three dimensions especially,  
the missed part may turn out to be a set of measure zero---a starting ansatz of this approach. 
$\hns$ derived below seems to be written with the natural dynamical coordinates 
for studying fluid particle separations over time. 
Future papers will study this further with the classical similarity renormalization group and 
dissipative canonical Poisson bracket structure derived in this paper. 

\subsection{Nonholonomic constraint for Navier-Stokes Hamiltonian}

The equations of motion follow from Hamilton's principle that the variation of an action vanishes:
\be
\del\cs=\int d^4x\,\del\cl=\int dt\,d^3x\,\del\cl=0\;,
\ee
where $\cl$ is a lagrangian density and the integral is over all space 
and an interval of time between  
arbitrary endpoint initial and final times for the dynamics of interest. Interactions of interest 
are simply added to $\cl$.\footnote{with the usual minus sign: 
$L=T-V$.}
This includes constraints of interest such as the entropy constraint 
of Eq.~(\ref{eq:scon}). Requiring it to be satisfied over all space gives, with some exact rearrangements,  
\bea
0&=&\int d^3x\left(\rho T D_t s-\sig_{ij}^\pr\pj v_i+\pj q_j\right)\nn\\
&=&\int d^3x\left[\rho T \pt s+\rho T v_i\pdi s+v_i\,\pj\sig_{ij}^\pr 
+\pj (q_j-v_i\,\sig_{ij}^\pr)\right]\nn\\
&=&\int d^3x\left[\rho T \pt s+\rho T v_i\pdi s+v_i\,\pj\sig_{ij}^\pr\right] 
+\int d^2x\,\hat{n}_j\left(q_j-v_i\,\sig_{ij}^\pr\right)\;,\nn
\eea
using the divergence theorem once again to convert the volume integral into a surface integral 
with unit outward normal $\bnh$. Note that the integrand of this surface integral in the  
last term is the 
exact same irreversible energy flux discussed below the energy conservation equation, Eq.~(\ref{eq:energy}). 
Since heat flux $\bq$ is proportional to temperature (and in general chemical potential) gradients and $\sig^\pr_{ij}$ is 
proportional to velocity gradients, 
and all field gradients are assumed to vanish at spatial infinity, this term can be safely dropped. Thus 
the constraint to be satisfied becomes 
\be
0=\int d^3x\left[\rho T \pt s+v_i\left(\rho T\pdi s+\pj\sig_{ij}^\pr\right)\right]\;,\label{eq:scon2}
\ee
where we have rearranged it for convenience as described next. 
Eq.~(\ref{eq:scon2}) is a nonholonomic path-dependent constraint (with pathlines given by $v_i$) \cite{hiroki}
due to the irreversible dynamics of heat being added to the volume through viscosity. 
Thus the Euler Lagrangian density, $\cle$ of Eq.~(\ref{eq:leuler}), can not be simply modified by 
changing its last ideal entropy constraint term to the above. 
Rather we must introduce initial-position vector field $\bax$ \cite{hiroki} as discussed in the 
opening paragraph of this section. 

Nonholonomic constraints such as Eq.~(\ref{eq:scon2}) must be expressed 
as ``a linear relation connecting the {\it differentials} of the [coordinates]'' \cite{goldstein}. 
In field theory the meaning of coordinate 
is necessarily generalized to coordinate {\it field}. 
Thus it is natural to introduce initial-position vector field $\bax$, which is just a coordinate transformation, 
and to seek 
a linear relation connecting its variations, $\del\bax$, with those of the other 
coordinate field pertinent to any discussion of heat dissipation: 
specific entropy variation $\del s(\bx,t)$. 
As detailed in \cite{hiroki, zakharov}, $\bax$ denotes the initial position of  
the fluid particle that happens to be at $\bx$ at time $t$. 
At initial time $t_0$ we have 
\be
\ba(\bx,t_0)=\bx
\;.
\ee
In other words, $\ba$ is just another $\bx$---but it is the initial one 
and an assumed one-to-one map to all of the fluid particles. 
Consult \cite{zakharov} for further discussions including 
the relabeling (gauge) symmetry of fluid particles that $\ba$ satisfies. 
So $\bax$ is just a coordinate transformation and we assume that it is a  
non-singular one, i.e.\ its Jacobian never vanishes:
\be
J\equiv \frac{\pd(A_1,A_2,A_3)}{\pd(x_1,x_2,x_3)}\equiv\det(\hat{J}_{ij})\equiv\det\left(\frac{\pd A_i}{\pd 
x_j}\right)\equiv\det\left(\pj A_i\right)\neq 0\label{eq:assump1}
\;,
\ee 
where like \cite{zakharov} the hat on $\hat{J}$ implies ``matrix'' and 
$\pdi=\pd/\pd x_i$ is the same shorthand that we have been using throughout this 
paper.\footnote{Although ``$\ba$ is just another $\bx$,'' to manage our notation and in keeping with the 
Euler representation of a fluid, we will always write out the derivatives with respect to $\ba$ 
in full such as $\pd/\pd A_i$; but whenever we come across 
$\pd/\pd x_i$ we will continue to replace it by the convenient $\pdi$. Further motivation for a 
distinction in the notation  
is that $\pd/\pd A_i = \hat{J}^{-1}_{ji}\pdj$, so it is a nonlocal operator---whereas $\pdi$ is local by definition.}

Although this is not the most general choice \cite{sudarshan}, it is nevertheless intuitive 
and leads to the standard Navier-Stokes dynamics as shown below; therefore, we  
assume that $\ba$ is conserved along all fluid path lines:
\be
\fbox{$\pt \ba+(\bv\cdot\bn)\ba=D_t\ba=0$}\label{eq:assump2}
\;.
\ee
This is a very important dynamical assumption to what follows: 
{\it Through it deriving the Navier-Stokes Hamiltonian becomes soluble}. Given Eq.~(\ref{eq:assump1}), 
inverting Eq.~(\ref{eq:assump2}) for the pathlines, $v_i$, is easy and follows from the chain rule of  
partial differentiation: 
\be
\fbox{$
v_i=-\frac{\pd x_i}{\pd A_j}\pt A_j$}\label{eq:viminus}
\;;
\ee
just multiply this with $\pdi A_k=\frac{\pd A_k}{\pd x_i}$, use $\frac{\pd A_k}{\pd A_j}=\del_{jk}$,
and it gives back Eq.~(\ref{eq:assump2}).
Substituting Eq.~(\ref{eq:viminus}) into Eq.~(\ref{eq:scon2}) changes the nonholonomic constraint to
\be
0=\int d^3x\left[\rho T \pt s-\left(\rho T\pdi s+
\pk\sig_{ik}^\pr\right)\left(\frac{\pd x_i}{\pd A_j}\right)\pt A_j\right]\;.\label{eq:scon3}
\ee
Now, just like was done with some of the thermodynamic differential relations earlier, 
these partial time derivatives in Eq.~(\ref{eq:scon3}) can be generalized to arbitrary field variations. 
The point is that the two field variations ($\del\ba$ and $\del s$) are proportional to each other and they can be either 
spatial or temporal or general variations as given by a differentiable manifold. 
Thus the final form for the nonholonomic constraint, which is the 
same entropy constraint as Eq.~(\ref{eq:scon}) rewritten for the $\hns$ derivation, is \cite{hiroki}
\be
\fbox{$
\rho T \del s=\left(\rho T\pdi s+
\pk\sig_{ik}^\pr\right)\left(\frac{\pd x_i}{\pd A_j}\right)\del A_j$}\;.\label{finalcons}
\ee
This is the relation of Goldstein's \cite{goldstein} that we were seeking between the coordinate field variations.
It may be instructive to note that this factor $\frac{\pd x_i}{\pd A_j}$ is just 
the matrix inverse of the gradient of vector field $\ba$:
\be
\left(\frac{\pd x_i}{\pd A_j}\right)\bn_k A_j=
\frac{\pd x_i}{\pd A_j}\hat{J}_{jk}=
\frac{\pd x_i}{\pd A_j}\frac{\pd A_j}{\pd x_k}=
\frac{\pd x_i}{\pd x_k}=\del_{ik}\;.
\ee
In other words: $\frac{\pd x_i}{\pd A_j}=\hat{J}_{ij}^{-1}$. Thus, $\frac{\pd x_i}{\pd A_j}$ 
in entropy constraint Eq.~(\ref{finalcons}) is a nonlocal object 
with derivatives of vector field $\ba$ in the denominator.

Before leaving this section we discuss mass conservation in these new coordinates. 
Mass conservation equivalent to Eq.~(\ref{eq:1}) is now simply given by the following 
kinematic constraint~\cite{hiroki}
\be
\fbox{$
\rho(\bx,t)=J\,\rho_0=J(\bn\ba)\,\rho_0(\ba)$}\;,\label{eq:mass2}
\ee
where $J$ is the Jacobian of 
Eq.~(\ref{eq:assump1}) and all of the field dependencies are 
written out on the right for clarity.\footnote{Notation warning: 
what is written here as $J$ for the Jacobian is written as 
its inverse $J^{-1}=1/J$ (and vice versa) in both \cite{zakharov} and \cite{hiroki}. We find 
the notation $J(\bn\ba)$ convenient for explicitly showing the local Jacobian field dependencies.} 
As shown below in the Navier-Stokes equation section, taking a partial time derivative 
of Eq.~(\ref{eq:mass2}) exactly reproduces the continuity equation, Eq.~(\ref{eq:1}), as an operator and thus they are 
equivalent, but here we close this section with a simpler proof. 
Integrating Eq.~(\ref{eq:mass2}) over all space and 
using the usual rules of a Jacobian shows that the total mass of a fluid:
\be
\intx\,\rho(\bx,t)= \intx J \rho_0=\int d^3\!A\,\rho_0(\ba)
\;,
\ee
is the same in both coordinate systems, i.e.\ mass is conserved. Also, the inverse Jacobian 
is the volume strain of the fluid: $J^{-1}\sim\,d^3x/d^3A$. Now we have all the pieces for $H_{NS}$ 
and we move on to its derivation.

\subsection{Navier-Stokes Lagrangian}

Just like in the Euler Hamiltonian section above, the derivation of the canonical 
Navier-Stokes Hamiltonian, $H_{NS}$, 
begins with a definition of its lagrangian density. As motivated by earlier discussions, 
this time we use the form of 
Fukagawa and Fujitani \cite{hiroki}:
\be
\fbox{$
\clns\left(\rho,\bv,s,K,\ba,\bb\right)=\onehalf \rho\bv^2-\cu(\rho,s)+
\bb\cdot D_t\ba+K\left[\rho-J(\bn\ba)\,\rho_0(\ba)\right]
$}\;.\label{eq:lns}
\ee
Most of the variables have been defined earlier, however vector field $\bb$ and scalar field $K$ are new. 
They are simply the lagrange multiplier fields for the respective constraints of 
$\ba$ being conserved with the flow and mass be conserved generally. 
Vector field $\bb$ also ends up being the conjugate momentum field to $\ba$ itself 
(as our notation anticipates), 
so these two fields together end up being the only dynamical fields of $\clns$ and 
consequently $\chns$ as well.  
Finally, it is important to note that $\clns$ is implied to be augmented with the nonholonomic constraint 
of Eq.~(\ref{finalcons}) that relates the $\del s(\bx,t)$ variations to those of $\del \ba(\bx,t)$. 
Actually, as emphasized by \cite{hiroki}, because of dissipation, $\clns$ (and therefore $\chns$ too) is not  
a function of the specific entropy field $s(\bx,t)$, but rather can only be specified in terms of its 
variations. This is a consequence of $s(\bx,t)$ satisfying a nonholonomic constraint. 
This causes no problems in principle because the general dynamics of a theory come from 
the Hamilton equations and Poisson brackets (which evolve any other physical observables 
of interest in time or scale) and these follow 
from {\it variations} of $H_{NS}$ and the observables as shown below.   

Now we move on to the Euler-Lagrange equations for $\clns$ and show that they are 
equivalent to the Navier-Stokes equation (with mass and energy conservation).  
Then $\hns$ is derived and its Hamilton equations are shown 
to be these same Euler-Lagrange equations.  

\subsection{Euler-Lagrange equations for viscous fluid}
\label{sec:ELeq}

The Euler-Lagrange equations of $\clns$ with nonholonomic constraint Eq.~(\ref{finalcons}) 
follow straightforwardly from Hamilton's principle as shown next. 
To start, note that the following fields of $\clns$ all have a simple Euler-Lagrange equation 
(with no spatial or temporal partial derivatives of the respective fields in $\clns$): 
$\bv$, $\bb$, $\rho$, and $K$. Thus for simplicity of notation we define the following temporary collection of fields 
by
\be
\chi_\alpha = (\bv, \bb, \rho, K)\label{eq:sh}
\ee
for $\alpha=1,\ldots,8$; 
this is just a shorthand for $\chi_1 = v_1$, $\chi_2 = v_2$, $\ldots$, $\chi_{8} = K$. 
Then Hamilton's principle on the Navier-Stokes action with constraint Eq.~(\ref{finalcons}) becomes 
\bea
&&\!\!\!\!\!\!\!\!0=\del{\cal S}_{NS}=\int d^4x\,\del\clns=\nn\\
&&\!\!\!\!\!\!\!\!\!\int d^4x\,\left[\frac{\pd\clns}{\pd \chi_\alpha}\delta \chi_\alpha
+\frac{\pd\clns}{\pd s}\frac{\del s}{\del A_i}\delta A_i
+\frac{\pd\clns}{\pd A_i}\delta A_i
+\frac{\pd\clns}{\pd (\pt A_i)}\delta (\pt A_i)
+\frac{\pd\clns}{\pd (\pj A_i)}\delta (\pj A_i)\right]\label{eq:lns2}
\eea
The integrand of the second term from Eqs.~(\ref{eq:tdef}) and (\ref{eq:lns}) becomes 
\be
\del\cl^{(s)}_{NS}\equiv\frac{\pd\clns}{\pd s}\frac{\del s}{\del A_i}\delta A_i=
-\frac{\pd\cu}{\pd s}\frac{\del s}{\del A_i}\delta A_i=
-\rho T\frac{\del s}{\del A_i}\delta A_i\;,
\ee
which will often be left in this form for simplicity; however it is implied by this last expression that 
the remaining variational derivative is to be replaced by the nonholonomic constraint of Eq.~(\ref{finalcons}) leaving 
\be
\frac{\del\cl^{(s)}_{NS}}{\delta A_i}=\fbox{$-\rho T\frac{\del s}{\del A_i}=
-\left(\rho T\pj s+\pk\sig_{jk}^\pr\right)\frac{\pd x_j}{\pd A_i}$}\label{eq:22}
\;.
\ee
This term comes up often in what follows. 
One way to state the results of varying $\clns$ is to say that  
all of its functional derivatives follow the 
standard procedures of field theory except for the ones with respect to 
vector field $\ba$: because of dissipation, these variations, 
$\del\cl_{NS}/\del A_i$, are appended with this boxed term, $\del\cl^{(s)}_{NS}/\delta A_i$.

We continue with a few standard looking steps deriving the Euler-Lagrange equations 
that result from Eq.~(\ref{eq:lns2}) 
because it will be useful in the Hamiltonian and Poisson bracket sections later, and also   
to emphasize that these steps would not follow so simply except for the fact that 
coordinate transformation $\ba(\bx,t)$ has been introduced. As emphasized by \cite{hiroki}, 
the spatial and temporal 
boundary conditions of the surface terms could not be met if it were not for 
the specific entropy variations $\del s$ being replaced by those of $\del\ba$. 
Handling $\del\cl^{(s)}_{NS}$ as above, 
continuing with Eq.~(\ref{eq:lns2}) in the standard way 
gives two surface terms and two sign flips:
\bea
&&\!\!\!\!\!\!\!\!0=\del{\cal S}_{NS}=\int d^4x\,\del\clns=\nn\\
&&\!\!\!\!\!\!\!\!\!\int d^3x\left.\frac{\pd\clns}{\pd (\pt A_i)}\delta A_i\right|_{{\rm temporal}~{\rm endpoints}}
+\int dt\,d^2x\,\hat{n}_j\left.\frac{\pd\clns}{\pd (\pj A_i)}\delta A_i\right|_{{\rm spatial}~{\rm boundary}}+\nn\\
&&\!\!\!\!\!\!\!\!\!\int d^4x\,\left\{\frac{\pd\clns}{\pd \chi_\alpha}\delta \chi_\alpha+
\left[\frac{\pd\clns}{\pd A_i}
-\pt\left(\frac{\pd\clns}{\pd (\pt A_i)}\right)
-\pj\left(\frac{\pd\clns}{\pd (\pj A_i)}\right)
-\rho T\frac{\del s}{\del A_i}\right]\delta A_i\right\}\;.\label{eq:lns3}
\eea
Since $\ba$ has been introduced, 
these surface terms vanish by the same physical ansatz used for the ideal fluid: that 
coordinate field variations vanish at spatial infinity and the temporal endpoints. 
$\del s(\bx,t)$ cannot be made to vanish like this, its variations depend on the path taken \cite{hiroki}. 
The remaining fields are independent and their variations are arbitrary, 
thus this implies the following Euler-Lagrange 
equations for the viscous fluid:
\bea
\frac{\pd\clns}{\pd \chi_\alpha}&=&0\;,\label{EL1}\\
\pt\left(\frac{\pd\clns}{\pd (\pt A_i)}\right)&=&\frac{\pd\clns}{\pd A_i}
-\pj\left(\frac{\pd\clns}{\pd (\pj A_i)}\right)
-\rho T\frac{\del s}{\del A_i}\;,\label{EL2}
\eea
where recall $\chi_\alpha$ is just a convenient shorthand for the two vector and two scalar fields of Eq.~(\ref{eq:sh}).

Before carrying out these Euler-Lagrange equations, there are two math results to discuss. 
First, recall the definition of Jacobian $J$ discussed above and defined by Eq.~(\ref{eq:assump1}).
Note that it depends on the gradient of $\ba$ and that $\clns$ of Eq.~(\ref{eq:lns}) contains 
a factor of $J$ in the mass conservation constraint. Thus, for the Euler-Lagrange equations, we need 
to know how to take a derivative 
of this Jacobian with respect to $\pj A_i$. This follows from the so-called method of cofactors \cite{hiroki, dirac}:
\be
\frac{\pd J}{\pd(\pj A_i)}=J\,\frac{\pd x_j}{\pd A_i}=J\,\hat{J}_{ji}^{-1}\label{eq:result1}
\;.
\ee
Second, note that a further derivative of this result vanishes \cite{hiroki}:
\be
\pj\left(\frac{\pd J}{\pd(\pj A_i)}\right)=\pj\left(J\,\frac{\pd x_j}{\pd A_i}\right)=
\pj\left(J\,\hat{J}_{ji}^{-1}\right)
=0\label{eq:result2}
\;.
\ee
These two relations will be used in deriving the equations of motion below. The second of these equations, 
Eq.~(\ref{eq:result2}), is useful in canceling terms between the messes that arise.  
It follows from straight-forward algebra, with the partial derivatives of the two factors 
of the product exactly canceling, 
however to obtain the result note the following more general equation for the method of cofactors 
from Dirac's book again \cite{dirac}:
\be
\frac{\pd J}{\pd x_k}\equiv\pk J\equiv \pk\left[\det\left(\pj A_i\right)\right]=J\,\frac{\pd x_j}{\pd A_i}\pk(\pj A_i)
\label{eq:diracmagic}
\;.
\ee
At first sight the final result on the right looks a little unwieldy, but with all the factors explicitly shown 
in the determinant, it follows straightforwardly from the chain rule of calculus. To prove Eq.~(\ref{eq:result2}),
also note the following simple relations:
\beaa
&&~~~~\hat{J}\hat{J}^{-1}=\mathbb{1}\;,\\
&\Rightarrow&(\pd\hat{J})\hat{J}^{-1}+\hat{J}(\pd\hat{J}^{-1})=0\;,\\
&\Rightarrow&\pd\hat{J}^{-1}=-\hat{J}^{-1}(\pd\hat{J})\hat{J}^{-1}\;.
\eeaa

Back to the Euler-Lagrange equations: Varying the independent fields of $\clns$, 
according to Eqs.~(\ref{EL1}) and (\ref{EL2}), gives 

\underline{Varying $\bv$}:
\bea
&&~~~~~~~~~~\frac{\pd\clns}{\pd v_i}=0\;,\nn\\
&&\Rightarrow~~\rho\,v_i+\Pi_j \pdi A_j=0\;,\nn\\
&&\Rightarrow~~~\fbox{$\rho\,\bv=-\Pi_i \bn A_i$}\;.\label{eq:nsv}
\eea
This is the velocity constraint for the viscous fluid.
Note how written this way, in terms of momentum density $\rho\,\bv$, 
the right-hand side does not depend on density field $\rho(\bx,t)$. 
Once again, like with the velocity constraint for the ideal fluid, 
after these Euler-Lagrange equations have been derived, when
`$\rho\,\bv$' is written, `$-\Pi_i\bn A_i$' is implied.
Continuing with the remaining fields of $\clns$ gives

\underline{Varying $\bb$}:
\be
\Rightarrow~~\fbox{$D_t\ba=0$}\;;\label{eq:nsdta}
\ee 

\underline{Varying $\rho$}:
\bea
&&~~~~~~~~~~~~~\frac{\pd\clns}{\pd \rho}=0\;,\nn\\
&&~~\Rightarrow~~\frac{\bv^2}{2}-\frac{\pd\cu}{\pd\rho}+K=0\;,\nn\\
&&~~\stackrel{(\ref{eq:pdef})}{\Rightarrow}~~\frac{\bv^2}{2}-h+K=0\;,\nn\\
&&~~\Rightarrow~~~~~\fbox{$K=h-\frac{\bv^2}{2}$}\;;\label{eq:nsrho}
\eea

\underline{Varying $K$}:
\bea 
&&~~~~~\;\frac{\pd\clns}{\pd K}=0\;,\nn\\
&&\Rightarrow~~~\fbox{$\rho=J\,\rho_0$}\;\label{eq:masscon}
\eea
where $\rho_0(\ba(\bx,t))$ is the initial density, set by the physics of the problem, and 
$J$ is the Jacobian of Eq.~(\ref{eq:assump1}), a function of the $\ba$ gradient and given explicitly by \cite{schutz}
\bea 
J(\bn\ba)&\equiv& \frac{\pd(A_1,A_2,A_3)}{\pd(x_1,x_2,x_3)}\nn\\
&=&\eps_{ijk}\,(\pdi A_1)(\pj A_2)(\pk A_3)\nn\\
&=&\frac{1}{6}\eps_{ijk}\eps_{lmn}\,(\pdi A_l)(\pj A_m)(\pk A_n)\label{eq:6terms}\;,
\eea
with six terms in total; 
$\eps_{ijk}$ is the standard antisymmetric Levi-Civita symbol with $\eps_{123}\equiv~1$. 
Thus we see that Jacobian $J$ is a cubic derivative interaction which  is  quite complex. 
Below, we show how $J$ is essential to get these Euler-Lagrange  
equations of a viscous fluid to be exactly equivalent to 
the Navier-Stokes equation. However, note that the physics of the problem 
often allows density $\rho = J\rho_0$ to be approximated as a constant or near-constant starting point, 
thus giving $J(\bn\ba)\approx 1$ (when the density is approximately 
constant for all time). In the compressible (exact) case, as already mentioned, 
below we show how a partial time derivative of this mass constraint, $\rho = J\rho_0$, 
is exactly equivalent to the continuity equation. 

The final equation of motion for vector field $\ba$  
is the hard one, but it is straightforward given the above 
setup. Thus, the final Euler-Lagrange equation comes from  

\underline{Varying $\ba$ (and $s$ with implicit $\ba$ dependence)}:
\bea
\pt\left(\frac{\pd\clns}{\pd (\pt A_i)}\right)\!\!\!\!\!
&\stackrel{(\ref{EL2})}{=}&~~~~\;\frac{\pd\clns}{\pd A_i}
-\pj\left(\frac{\pd\clns}{\pd (\pj A_i)}\right)
-\rho T\frac{\del s}{\del A_i}\;,\nn\\
\Rightarrow~\pt \Pi_i&=&-K J \frac{\pd\rho_0}{\pd A_i}
-\pj\left(\Pi_i v_j-K\rho_0\frac{\pd J}{\pd (\pj A_i)}\right)
-\rho T\frac{\del s}{\del A_i}\nn\\
&\stackrel{(\ref{eq:result1})}{=}&-K J \frac{\pd\rho_0}{\pd A_i}
-\pj\left(\Pi_i v_j-K\rho_0\,J\,\frac{\pd x_j}{\pd A_i}\right)
-\rho T\frac{\del s}{\del A_i}\nn\\
&\stackrel{(\ref{eq:22})}{=}&-K J \frac{\pd\rho_0}{\pd A_i}
-\pj\left(\Pi_i v_j-K\rho_0\,J\,\frac{\pd x_j}{\pd A_i}\right)
-\left(\rho T\pj s+\pk\sig_{jk}^\pr\right)\frac{\pd x_j}{\pd A_i}\nn\\
&\stackrel{(\ref{eq:result2})}{=}&-K J \frac{\pd\rho_0}{\pd A_i}
-\pj\left(\Pi_i v_j\right)
+\pj\left(K\rho_0\right)J\,\frac{\pd x_j}{\pd A_i}
-\left(\rho T\pj s+\pk\sig_{jk}^\pr\right)\frac{\pd x_j}{\pd A_i}\nn\\
&\stackrel{({\rm chain}~{\rm rule})}{=}&
-\pj\left(\Pi_i v_j\right)
+(\pj K)\rho_0\,J\,\frac{\pd x_j}{\pd A_i}
-\left(\rho T\pj s+\pk\sig_{jk}^\pr\right)\frac{\pd x_j}{\pd A_i}\;,\nn\\
&\stackrel{(\ref{eq:masscon})}{\Rightarrow}&~\fbox{$\pt \Pi_i+\bn\cdot\left(\Pi_i\bv\right)=
\left[\rho\,\pj K-\rho\,T\pj s
-\pk\sig_{jk}^\pr\right]\frac{\pd x_j}{\pd A_i}$}\;.\label{eq:nsA}
\eea
Note that by using Eqs.~(\ref{eq:gradh}) and (\ref{eq:nsrho}) a perhaps more intuitive form of this last equation follows by 
noting the non-density pieces of the first two terms on the right are equal to the following:
\be
\bn K-T\bn s=\frac{\bn p}{\rho}-\bn\left(\frac{\bv^2}{2}\right)\nn
\;,
\ee
with a static and dynamic pressure contribution to this acceleration. Thus, this equation of motion for $\bb$ 
can be rewritten as
\bee
\fbox{$\pt \Pi_i+\bn\cdot\left(\Pi_i\bv\right)=
-\left[\rho\,\pj\left(\frac{\bv^2}{2}\right)+\pk\sig_{jk}\right]\frac{\pd x_j}{\pd A_i}$}\;.
\eee
Now this is looking like the Navier-Stokes equation, Eq.~(\ref{eq:2b}), with the full 
stress tensor term; in the next section, we show that they are exactly equivalent.

All of the equations of motion of $\clns$ have now been derived and we move on to the proof 
that they are equivalent to the Navier-Stokes and mass conservation equations 
(along with energy conservation which is maintained as an entropy constraint in this approach). 

\subsection{Navier-Stokes equation equivalence proofs}
\label{sec:nseep}

Here we show that the equations of motion of $\clns$ derived in the last section 
are equivalent to the mass, momentum, and energy conservation equations of Section~\ref{sec:conservation}. 
First, note that mass and energy conservation 
are automatically maintained via kinematic constraints in this approach defined by    
lagrangian density $\clns$ of Eq.~(\ref{eq:lns}) and 
entropy constraint Eq.~(\ref{finalcons}). Thus, 
one way to complete the proof is the following:
(i) show that $\rho=J\rho_0$ is equivalent to the continuity equation, Eq.~(\ref{eq:1}), and 
(ii) show that the Navier-Stokes momentum conservation equation, Eq.~(\ref{eq:2}), is equivalent to the Euler-Lagrange equations 
of $\clns$ from the previous section. We perform these two steps in order just below, but first for reference, here 
we list all of the equations of motion that came from these Euler-Lagrange equations of $\clns$
(including the entropy constraint for completeness): 
\bse
\label{eq:wow}
\bea
\frac{\del s}{\del A_i}~&=&~\left(\pj s+\frac{\pk\sig_{jk}^\pr}{\rho T}\right)\frac{\pd x_j}{\pd A_i}\;,\label{eq:000}\\
\rho~&=&~J\,\rho_0\;,            \label{eq:333}\\
\rho\bv~&=&~-\Pi_i\bn A_i\;,       \label{eq:111}\\
K~&=&~h-\frac{\bv^2}{2}\;,       \label{eq:222}\\
D_t\ba~&=&~0\;,                  \label{eq:444}\\
\pt \Pi_i+\bn\cdot\left(\Pi_i\bv\right)~&=&~\left[\rho\,\pj K
-\rho\,T\pj s-\pk\sig_{jk}^\pr\right]\frac{\pd x_j}{\pd A_i}\;.\label{eq:555}
\eea
\ese
The $\ba$ and $\bb$ equations are dynamical---the remaining equations are kinematic constraints. 
Next we show that these equations contain the same dynamics as that of  
the Navier-Stokes equation. 

\subsubsection{Mass conservation equivalence proof for viscous fluid}

Here we take a partial time derivative of the so-called mass operator: 
\be
\com\equiv\rho-J\,\rho_0\;, 
\ee 
and show that it exactly reproduces the continuity equation, Eq.~(\ref{eq:1}). 
Using the previously derived relations as referenced here above the equal signs, 
taking this time partial using the product rule of differentiation gives
\bea
\fbox{$\pt\com$}&=&\pt\rho-\rho_0\,(\pt J)-J\,(\pt \rho_0)\nn\\
&\stackrel{({\rm temporal}~(\ref{eq:diracmagic}))}{=}&\pt\rho
-\rho_0\left(J\,\frac{\pd x_j}{\pd A_i}\pt(\pj A_i)\right)-J\,(\pt \rho_0)\nn\\
&\stackrel{({\rm chain}~{\rm rule})}{=}&\pt\rho
-\rho_0\left(J\,\frac{\pd x_j}{\pd A_i}\pj(\pt A_i)\right)-J\,\frac{\pd\rho_0}{\pd x_j}\frac{\pd x_j}{\pd A_i}\pt A_i\nn\\
&\stackrel{(\ref{eq:result2})}{=}&\pt\rho
-\pj\left(\rho_0\,J\,\frac{\pd x_j}{\pd A_i}\pt A_i\right)\nn\\
&\stackrel{(\ref{eq:333})}{=}&\pt\rho
-\pj\left(\rho\,\frac{\pd x_j}{\pd A_i}\pt A_i\right)\nn\\
&\stackrel{(\ref{eq:viminus})}{=}&\fbox{$\pt\rho+\pj\left(\rho v_j\right)$}\;,
\eea
with the sign flipping exactly as required in this last line according to Eq.~(\ref{eq:viminus}). 
Thus the operators are exactly equivalent, and if one vanishes then so must the other: mass is conserved, q.e.d.

\subsubsection{Momentum conservation equivalence proof for viscous fluid}

Start with the Navier-Stokes momentum operator of Eq.~(\ref{eq:2}) with all the terms on the left-hand side:
\bea 
\cons&\equiv&\pt(\rho v_i)+\pj(\rho v_iv_j)-\pj\sig_{ij}\nn\\
&=&v_i\left[\pt\rho+\bn\cdot(\rho\bv)\right]+\rho\left[\pt v_i+v_j\pj v_i-\frac{\pj\sig_{ij}}{\rho}\right]
\;,
\eea
then insert the velocity constraint of Eq.~(\ref{eq:111}) and the thermodynamic law of Eq.~(\ref{eq:gradh}), 
and finally exactly rearrange to show that $\cons$ is proportional to the Euler-Lagrange equations of motion of 
Eq.~(\ref{eq:wow}). 

First, we need one preliminary math result. The equation of motion for $\bb$, Eq.~(\ref{eq:555}), 
has a matrix $\pd x_j/\pd A_i$ on the right-hand side that that needs to be inverted. 
Therefore, multiplying  Eq.~(\ref{eq:555}) by $\pd A_i/\pd x_m$, performing the implied sum over $i$, using the chain rule, 
and rearranging dummy indices, produces an equivalent form of the equation of motion of $\bb$:
\be
\left[\pt \Pi_j+\bn\cdot\left(\Pi_j\bv\right)\right]\pdi A_j~=~\rho\,\pdi K
-\rho\,T\pdi s-\pj\sig_{ij}^\pr\;.\label{eq:555b}
\ee
We mention this because this is the form that appears in the proof below and this shows that it is equivalent 
to Eq.~(\ref{eq:555}).

Recall Eq.~(\ref{eq:3}) for the relation between the stress tensor and the viscous stress tensor. 
Then starting with $\cons$ above, insert Eqs.~(\ref{eq:111}) and (\ref{eq:gradh}). Then use the 
product rule and rearrange the operator adding and subtracting the same term in two spots, and massage it 
into its final form proportional to the Euler-Lagrange equations as shown here: 
\bea
\cons&\equiv&\pt(\rho v_i)+\pj(\rho v_iv_j)-\pj\sig_{ij}\nn\\
&=&\pt\left[-\Pi_j\pdi A_j\right]+\pk\left[-\Pi_j(\pdi A_j)v_k\right]
+\rho\pdi h-\rho T\pdi s-\pj\sig^\pr_{ij}\nn\\
&=&\rho\pdi h-\rho T\pdi s-\pj\sig^\pr_{ij}-(\pt \Pi_j)\pdi A_j-\Pi_j\pdi (\pt A_j)\nn\\
&&-(\pk \Pi_j)(\pdi A_j)v_k-\Pi_j(\pdi\pk A_j)v_k-\Pi_j(\pdi A_j)\pk v_k\nn\\
&\stackrel{({\rm exact}~{\rm rearrange})}{=}&
\rho\pdi h-\rho T\pdi s-\pj\sig^\pr_{ij}
-(\pdi A_j)\left[\pt \Pi_j+\pk(\Pi_jv_k)\right]\nn\\
&&-\Pi_j\pdi\left(D_tA_j\right)+\Pi_j(\pdi v_k)(\pk A_j)\nn\\
&\stackrel{(+/-\,\rho\,\pdi K)}{=}&
-\Pi_j\pdi\left\{D_tA_j\right\}\nn\\
&&+\left\{\rho\pdi K-\rho T\pdi s-\pj\sig^\pr_{ij}-(\pdi A_j)\left[\pt \Pi_j+\pk(\Pi_jv_k)\right]\right\}\nn\\
&&-\rho\pdi K+\rho\pdi h+\Pi_j(\pdi v_k)(\pk A_j)\nn\\
&\stackrel{(+/-\,\rho v_j\pdi v_j)}{=}&
-\Pi_j\pdi\left\{D_tA_j\right\}\nn\\
&&+\left\{\rho\pdi K-\rho T\pdi s-\pj\sig^\pr_{ij}-(\pdi A_j)\left[\pt \Pi_j+\pk(\Pi_jv_k)\right]\right\}\nn\\
&&+\rho\,\pdi\left\{h-K-\frac{\bv^2}{2}\right\}\nn\\
&&+(\pdi v_j)\left\{\rho v_j+\Pi_k\pj A_k\right\}\;.\label{eq:BIG}
\eea
The last four lines are the form of the equation that we were seeking. Each factor in 
curly braces is 
the Euler-Lagrange equation operator for the viscous fluid that we were seeking. They are the equations 
of motion for respective fields \\
$~~~~~~$(1) dynamical $\ba$, Eq.~(\ref{eq:444}); \\
$~~~~~~$(2) dynamical $\bb$, Eq.~(\ref{eq:555b}) (equiv.\ to Eq.~(\ref{eq:555}));\\
$~~~~~~$(3) constrained $K$, Eq.~(\ref{eq:222}); and\\
$~~~~~~$(4) constrained $\bv$, Eq.~(\ref{eq:111}).\\
Finally, the entropy constraint Eq.~(\ref{eq:000}) is built in and maintained for these Euler-Lagrange equations, 
and from the previous section, the mass constraint Eq.~(\ref{eq:333}) is independently equivalent to 
the continuity equation, Eq.~(\ref{eq:1}). Therefore it is all consistent and 
the Euler-Lagrange equations of $\clns$ are exactly equivalent to 
the Navier-Stokes equation, mass conservation, and energy conservation (or the entropy constraint 
however one wants to state it as discussed in Section~\ref{sec:conservation}). q.e.d.\
Interestingly, unlike for the ideal fluid (cf. Eq.~(\ref{eq:magic})), here for the viscous fluid,  
mass and momentum conservation decouple; so for example
violation of the mass constraint equation, $\rho=J\rho_0$, 
only affects mass conservation, but does not directly affect 
the dynamical Euler-Lagrange equations as shown by it not being present in Eq.~(\ref{eq:BIG}) above---which  
came from the momentum conservation operator alone. But with the ideal fluid, mass and momentum conservation are 
coupled as given by the final form of Eq.~(\ref{eq:magic}) which contains the continuity equation operator 
as its first term on the right-hand side. 
The difference lies in the fact that with the ideal fluid, $\rho$ is a dynamical field, 
but with the viscous fluid, $\rho$ is constrained (as appears to be required in order 
to introduce dissipation consistently). 

\subsection{Hamilton equations for viscous fluid} 
\label{sec:hameq}

Now we continue with the Navier-Stokes Hamiltonian 
derivation to express the problem in terms of its canonical coordinates so as 
to see its flow through phase space, and then as a function of scale with the SRG (see Appendix~\ref{app:csrg}). 
In a sense the hard part of this paper is complete: 
the Euler-Lagrange equations of a viscous fluid 
have been shown to be equivalent to the Navier-Stokes equation. 
Now, we put the remaining pieces together setting 
up the field theory from a Hamiltonian point of view,   
deriving its dissipative canonical Poisson bracket structure, and working out the 10-generator 
galilean group algebra of a nonrelativistic viscous fluid (see Appendix~\ref{app:sym}). 
The Hamilton equations of $\chns$ give rise to the same dynamical equations as those that came 
from the Euler-Lagrange equations of $\clns$. 
However, just like earlier in deriving the Euler-Lagrange equations of $\clns$, 
here too the functional derivative of $\bax$ has an extra term to the standard ones of field theory. 
Thus, for clarity, now we start from first principles and derive the Hamilton equations with dissipation. 

As before, the derivation of a Hamiltonian starts with a discussion of the conjugate momenta. 
For lagrangian density $\clns$ of Eq.~(\ref{eq:lns}) the nonzero conjugate momentum fields are
\be
\fbox{$
\pi_{A_i}\equiv\frac{\pd\clns}{\pd(\pt A_i)}=\Pi_i(\bx,t)$}\;,\label{eq:bday1}
\ee
where $i=(1,2,3)$ as always in this paper with three spatial dimensions 
(recall $\ba$ is a position vector of the same type as $\bx$); so this is really 
three independent conjugate momentum fields. 

At this point, because of the complexity that dissipation adds, we go through the 
derivation of the Hamilton equations which necessarily involves the definition 
of the Hamiltonian itself. Just like in Landau and Lifshitz's jewel \cite{landau2} 
but for field theory, start back with the expression used in deriving the Euler-Lagrange 
equations, but this time rearrange things so as to make the conjugate momenta come to the 
forefront via a so-called Legendre transformation. So back up to Eq.~(\ref{eq:lns2}), drop the spatial 
surface terms like before (which necessarily required us to introduce vector field $\bax$), 
and put in the definition of the conjugate momenta from Eq.~(\ref{eq:bday1}). This gives 
\be
\del\clns=\frac{\pd\clns}{\pd \chi_\alpha}\delta \chi_\alpha+\piai\del\left(\pt A_i\right)+
\left[\frac{\pd\clns}{\pd A_i}
-\pj\left(\frac{\pd\clns}{\pd (\pj A_i)}\right)
-\rho T\frac{\del s}{\del A_i}\right]\delta A_i\;,\label{eq:bday2}
\ee
where recall $\chi_\alpha$ is defined as in Eq.~(\ref{eq:sh})---it is just a convenient shorthand 
because all of its fields do not have a conjugate momentum field and so their Euler-Lagrange 
equations are all the simple `$\pd{\cal L}/\pd\chi_\alpha =0$'. 
Now, at this point the Euler-Lagrange equations of Eqs.~(\ref{EL1}) and 
(\ref{EL2}) are enacted\footnote{This is the point where the temporal endpoint boundary term is 
also dropped since that ansatz is used in deriving the Euler-Lagrange equations themselves.} 
and the following identification is made:
\be
\pt\left(\frac{\pd\clns}{\pd (\pt A_i)}\right)\rightarrow \pt\pi_{A_i} \;,
\ee
which follows simply from definition Eq.~(\ref{eq:bday1}) itself. Thus, Eq.~(\ref{eq:bday2}) becomes 
\be
\del\clns=\dot{\pi}_{A_i}\del A_i+\piai\del\dot{A}_i
\;,
\ee
where for cleanness of notation we use the dot notation for the temporal partial derivative here. 
This is the point where we leave the Lagrangian framework and move over to the Hamiltonian one: 
rearranging terms gives simply
\be
\del\left(\piai\dot{A}_i-\clns\right)=-\dot{\pi}_{A_i}\del A_i+\dot{A}_i\del\piai
\;.
\ee
Define the variational argument of the left-hand side as the hamiltonian density itself:
\be\fbox{$
\chns\equiv\piai\dot{A}_i-\clns$}\;,\label{eq:hnsdef}
\ee
and this variation becomes
\be
\del\chns=-\dot{\pi}_{A_i}\del A_i+\dot{A}_i\del\piai\;.\label{eq:90}
\ee
This implies the following Hamilton equations for the viscous fluid:
\bse
\label{eq:hameqns}
\bea
\dot{A}_i&=&\frac{\del\chns}{\del\piai}\;,\\
\dot{\pi}_{A_i}&=&-\frac{\del\chns}{\del A_i}\;.
\eea
\ese
We will derive these equations more explicitly just below, but first we 
write the Hamiltonian that Eq.~(\ref{eq:hnsdef}) implies. Thus the Naiver-Stokes 
Hamiltonian is given by
\bea
\hns&\equiv&\intx\,\chns\;,\label{eq:hnsint}\\
\chns&=&\piai\dot{A}_i-\clns\nn\\
&\stackrel{(\ref{eq:bday1})}{=}&\Pi_i\,\dot{A}_i-\clns\nn\\
&\stackrel{(\ref{eq:lns})}{=}&-\onehalf\rho\bv^2+\cu(\rho,s)
-\Pi_iv_j\pj A_i-K\left(\rho-J\,\rho_0\right)\;,
\label{eq:hns1}
\eea
which is not the final form of $\chns$ because there are constraints. 
In this derivation of $\chns$, the dynamical field equations for $\ba$ and $\bb$ are being rearranged, 
but the following constraints are assumed to always be upheld:
\bse
\label{eq:HnsCE}
\bea
\frac{\del s}{\del A_i}~&\stackrel{(\ref{eq:000})}{=}&~\left(\pj s
+\frac{\pk\sig_{jk}^\pr}{\rho T}\right)\frac{\pd x_j}{\pd A_i}\;,\label{eq:000b}\\
\rho~&\stackrel{(\ref{eq:333})}{=}&~J\,\rho_0\;,\label{eq:333b}\\
\rho\bv~&\stackrel{(\ref{eq:111})}{=}&~-\Pi_i\bn A_i\;,\label{eq:111b}\\
K~&\stackrel{(\ref{eq:222})}{=}&~h-\frac{\bv^2}{2}\;.\label{eq:222b}            
\eea
\ese

Thus, at first blush seemingly paradoxically (but its effects are still in there as shown below), 
the last term of Eq.~(\ref{eq:hns1}) vanishes and we are left with the following equation 
which with one slight rearrangement from the $\bv$ constraint leaves the final form 
of $\chns$:
\bea
\chns&=&-\onehalf\rho\bv^2+\cu(\rho,s)-\Pi_iv_j\pj A_i\nn\\
&\stackrel{(\ref{eq:111b})}{=}&-\onehalf\rho\bv^2+\cu(\rho,s)+v_j (\rho v_j)\nn\\
&=&\fbox{$\frac{(\rho\bv)^2}{2\rho}+\cu(\rho,s)$}_{~{\rm with}~{\rm constraints}~(\ref{eq:HnsCE})}\;.
\label{eq:hns0}
\eea
This is the Navier-Stokes Hamiltonian (density) that has been consistently derived 
for a nonrelativistic viscous fluid. Writing it in this final ``$P^2/(2M)$'' form is seen to be 
convenient when one recalls the form of the Navier-Stokes momentum density given by  
Eq.~(\ref{eq:111b}).\footnote{So this $(\rho\bv)^2=(\Pi_i\bn A_i)^2$ numerator actually does not depend on density field $\rho$, 
only the denominator does. Thus, the $\rho=J\rho_0$ in the denominator 
gives rise to nonlocal effects from the Jacobian $J$. See Eq.~(\ref{eq:6terms}) for 
explicit expressions for $J$.}

Now onto the dynamical equations that the Hamilton equations imply. First writing the hamiltonian density explicitly 
in terms of its dynamical fields $\ba$ and $\bb$ gives  
\be\fbox{$
\chns\left(\ba,\bb,s(\ba)\right)=\frac{(\Pi_i\bn A_i)^2}{2\,J(\bn\ba)\,\rho_0(\ba)}
+\cu(J(\bn\ba)\,\rho_0(\ba),s(\ba))$}\;.
\label{eq:theone}
\ee
This starts out quartic in the fields in the numerator of the first term and 
there is no standard quadratic term of field theory \cite{Lvov}. Recall that $J(\bn\ba)$ is the Jacobian explicitly 
given by Eq.~(\ref{eq:6terms}) with six terms in total and each term being 
cubic in $\pj A_i$---and for this first ``$P^2/(2M)$'' term it is in the denominator which
makes its contribution nonlocal. 
See the previous discussion after Eq.~(\ref{eq:6terms}) where possible approximate 
physical starting points are discussed regarding this Jacobian: e.g.\ $J(\bn\ba)\approx 1$ may be a 
natural starting point for an incompressible or near-incompressible fluid.

Now we start with $\chns$ of Eq.~(\ref{eq:theone}), finish the derivation of its Hamilton equations, and 
then show that they are equivalent to the Euler-Lagrange equations of the 
viscous fluid derived earlier and summarized by the six lines of 
Eq.~(\ref{eq:wow}) (which have already been shown to be equivalent to the Navier-Stokes equation itself 
supplemented with mass and energy conservation). 
Varying $\hns$ using the explicit form 
of Eq.~(\ref{eq:theone}) gives the following equation. 
Using thermodynamic relation Eq.~(\ref{eq:tdef}) and then integrating by parts leads to 
\bea
&&\!\!\!\!\!\!\!\!\del\hns=\intx\,\del\chns\left(\ba,\bb,s(\ba)\right)=\nn\\
&&\!\!\!\!\!\!\!\!\!\intx\,
\left[\frac{\pd\chns}{\pd A_i}\delta A_i
+\frac{\pd\chns}{\pd (\pj A_i)}\delta (\pj A_i)
+\frac{\pd\chns}{\pd s}\frac{\del s}{\del A_i}\delta A_i
+\frac{\pd\chns}{\pd \Pi_i}\delta \Pi_i
+\frac{\pd\chns}{\pd (\pj \Pi_i)}\delta (\pj \Pi_i)
\right]\nn\\
&&\!\!\!\!\!\!\!\!\!=\int d^2x\,\hat{n}_j
\left[
\frac{\pd\chns}{\pd (\pj A_i)}\delta A_i
+\frac{\pd\chns}{\pd (\pj \Pi_i)}\delta \Pi_i
\right]\nn\\
&&\!\!\!\!\!\!\!\!\!+\intx\,
\left\{\left[\frac{\pd\chns}{\pd A_i}
-\pj\left(\frac{\pd\chns}{\pd (\pj A_i)}\right)
+\rho\,T\frac{\del s}{\del A_i}\right]\delta A_i
+\left[\frac{\pd\chns}{\pd \Pi_i}
-\pj\left(\frac{\pd\chns}{\pd (\pj \Pi_i)}\right)\right]\delta \Pi_i
\right\}\label{eq97}
\eea
which is in terms of $\del\ba$ and $\del\bb$ variations alone as required in a hamiltonian treatment. 
The $\del s$ variations have been replaced 
with the correct $\del\ba$ ones given by nonholonomic constraint Eq.~(\ref{eq:000b}).
Dropping these spatial boundary terms on the physical grounds 
that $\del\ba$ and $\del\bb$ vanish at spatial infinity, and then   
comparing with the general form of $\del\chns$ already derived in Eq.~(\ref{eq:90}), gives the final operational form 
of the Hamilton equations for a viscous fluid:
\bse
\label{eq:bday3}
\bea
\fbox{$\pt A_i$}&=&~~\frac{\del\chns}{\del \Pi_i}=\fbox{$
\frac{\pd\chns}{\pd \Pi_i}
-\pj\left(\frac{\pd\chns}{\pd (\pj \Pi_i)}\right)$}
\;,\label{eq:itworks0}\\
\fbox{$\pt \Pi_i$}&=&-\frac{\del\chns}{\del A_i}=\fbox{$
-\frac{\pd\chns}{\pd A_i}
+\pj\left(\frac{\pd\chns}{\pd (\pj A_i)}\right)
-\rho\,T\frac{\del s}{\del A_i}$}\label{eq:itworks}
\;.
\eea
\ese
These followed straightforwardly even given dissipation. 

Now explicitly carrying out these derivatives of $\chns$ on the right-hand side of Eq.~(\ref{eq:bday3}),
we show that the Hamilton and Euler-Lagrange equations are equivalent for the viscous fluid; 
and that therefore as shown by Section \ref{sec:nseep}, they are also equivalent to the Navier-Stokes equation 
supplemented with mass and energy conservation. Clearly showing this was the main purpose of this paper. 

\subsubsection{Hamilton equations equivalence proof for viscous fluid}

The Hamilton and Euler-Lagrange equations are shown to be the same 
for the viscous fluid. Earlier it was shown that the Euler-Lagrange equations, summarized by  
Eq.~(\ref{eq:wow}), are equivalent to the Navier-Stokes equation. Thus this will complete 
the proof of showing that the Hamilton, Euler-Lagrange and conservation equations of 
the viscous fluid are all equivalent.

As discussed above, the Euler-Lagrange constraint equations are maintained in the Hamiltonian approach. 
So equivalence of the hamiltonian constraints, Eq.~(\ref{eq:HnsCE}), with those of Eq.~(\ref{eq:wow}) 
is already ``proven'' and only the 
dynamical Hamilton equations of $\ba$ and $\bb$ remain 
to be proven equivalent with their respective Euler-Lagrange equation. 

Thus, explicitly carrying out the derivatives of Eq.~(\ref{eq:bday3}) on Eq.~(\ref{eq:theone}) gives 
for $\ba$'s dynamical equation:
\bea
\pt A_i&=&\frac{\pd \chns}{\pd \Pi_i}-\pj\left(\,0\,\right)\nn\\
&=&\frac{\Pi_k(\pj A_k)(\pj A_i)}{J\rho_0}\nn\\
&\stackrel{(\ref{eq:333b})}{=}&\frac{\Pi_k(\pj A_k)(\pj A_i)}{\rho}\nn\\
&\stackrel{(\ref{eq:111b})}{=}&-\frac{\rho v_j(\pj A_i)}{\rho}\nn\\
&=&-v_j(\pj A_i)\nn\;,\\
\Rightarrow&\fbox{$D_t\ba=0$}&\;.
\eea

And now the hard one: the dynamical equation for $\bb$ as follows from Eq.~(\ref{eq:itworks}). 
One has to carefully take into account 
every $\ba$ dependent piece of Eq.~(\ref{eq:theone}), including 
the functional dependence of $\del s(\ba)$ on $\del\ba$; 
but note that this latter dependence has already been taken 
into account (as derived in Eq.~(\ref{eq97})) and is given by the final 
term on the right of the first line below. 
Thus, given the previous setup, 
$\bb$'s Hamilton equation follows straightforwardly from the product and chain rules of calculus:  
\bea
\pt \Pi_i&=&-\frac{\pd\chns}{\pd A_i}
+\pj\left(\frac{\pd\chns}{\pd (\pj A_i)}\right)
-\rho\,T\frac{\del s}{\del A_i}\nn\\
&\stackrel{(\ref{eq:theone})}{=}&
\frac{(\Pi_j\bn A_j)^2}{2\,J(\bn\ba)[\rho_0(\ba)]^2}\frac{\pd\rho_0}{\pd A_i}
-\frac{\pd\cu}{\pd\rho}\,J(\bn\ba)\,\frac{\pd\rho_0}{\pd A_i}-\rho\,T\frac{\del s}{\del A_i}\nn\\
&&+\pj\left(\frac{\Pi_i\Pi_k\pj A_k}{J(\bn\ba)\rho_0(\ba)}\right)
-\pj\left(\frac{(\Pi_k\bn A_k)^2}{2\,\rho_0(\ba)[J(\bn\ba)]^2}\frac{\pd J}{\pd(\pj A_i)}\right)\nn\\
&&+\pj\left(\frac{\pd\cu}{\pd\rho}\,\rho_0(\ba)\,\frac{\pd J}{\pd(\pj A_i)}\right)\nn\\
&\stackrel{(\ref{eq:pdef},\ref{eq:result1},\ref{eq:333b},\ref{eq:111b})}{=}&
\onehalf \bv^2J\frac{\pd\rho_0}{\pd A_i}
-h\,J\,\frac{\pd\rho_0}{\pd A_i}-\rho\,T\frac{\del s}{\del A_i}\nn\\
&&-\pj\left(\Pi_iv_j\right)
-\pj\left(\onehalf\rho_0\bv^2J\frac{\pd x_j}{\pd A_i}\right)
+\pj\left(h\rho_0J\frac{\pd x_j}{\pd A_i}\right)\nn\\
&\stackrel{(\ref{eq:result2},\,{\rm chain}\,{\rm rule})}{=}&
-\pj\left(\Pi_iv_j\right)
-\onehalf\left(\pj\bv^2\right)\rho\frac{\pd x_j}{\pd A_i}
+\left(\pj h\right)\rho\frac{\pd x_j}{\pd A_i}
-\rho\,T\frac{\del s}{\del A_i}\nn\\
&\stackrel{(\ref{eq:000b},\ref{eq:222b})}{=}&
-\pj\left(\Pi_iv_j\right)
+\left(\pj K\right)\rho\frac{\pd x_j}{\pd A_i}
-\left(\rho T\pj s+\pk\sig_{jk}^\pr\right)\frac{\pd x_j}{\pd A_i}\nn\;,\\
&\Rightarrow&\fbox{$\pt \Pi_i+\bn\cdot\left(\Pi_i\bv\right)=\left[\rho\,\pj K
-\rho\,T\pj s-\pk\sig_{jk}^\pr\right]\frac{\pd x_j}{\pd A_i}$}\label{100}
\eea
which is the same as Eq.~(\ref{eq:555}) as was to be shown. 
A lesson here is to note how the $\rho_0$ gradient and Jacobian derivative terms combined 
together in a beautiful way to produce this final exact result: {\it the $J$ and $\rho_0$ dependence 
of $\chns$ were essential for this exact result}. 
In summary, the Lagrangian, Eq.~(\ref{eq:lns}), 
and Hamiltonian, Eq.~(\ref{eq:theone}), both with nonholonomic entropy constraint, Eq.~(\ref{eq:000b}), 
satisfy the same equations of motion, those of a nonrelativistic viscous fluid 
given by the Navier-Stokes, continuity, and energy conservation equations of standard fluid dynamics. 
Also, if no approximations on any of the factors in $\clns$ or $\chns$ 
are made (such as numerical error in the process of integration etc.) then 
$\clns$ implies $\chns$ and vice versa. 
In summary, with mass constraint $\rho=J\,\rho_0$ and entropy constraint Eq.~(\ref{finalcons}) 
understood to be satisfied, as discussed amply above, a nice physical way to write the 
Navier-Stokes Hamiltonian is \cite{hiroki}
\be
\fbox{$
\hns=\intx\left[\frac{(\Pi_i\bn A_i)^2}{2\rho}+\cu(\rho,s)\right]$}\;.\label{eq:hns}
\ee

\subsection{Canonical Poisson brackets with dissipation}

We close the paper by showing that $dH_{NS}/dt=0$ even though $\hns$ includes dissipation, 
and then derive the Poisson bracket for a general classical dissipative observable of 
a nonrelativistic viscous fluid. First the time derivative of the Hamiltonian. Actually, 
the hard part here has already been done when deriving the Hamilton equations: 
from the general Navier-Stokes Hamiltonian variation derived earlier and 
given by Eq.~(\ref{eq:90}), combined with the explicit result of 
Eq.~(\ref{eq97}), the Hamilton equations and $\ba$ and $\bb$ functional 
derivatives followed as  
Eqs.~(\ref{eq:itworks0}) and (\ref{eq:itworks}). Thus the time derivative of $\hns$ is given by 
\bea
\frac{d\hns}{dt}&=&
\intx\left(\frac{\del\chns}{\del A_i}\frac{\pd A_i}{\pd t}+\frac{\del\chns}{\del \Pi_i}\frac{\pd \Pi_i}{\pd t}\right)\nn\\
&\stackrel{(\ref{eq:bday3})}{=}&
\intx\left(\frac{\del\chns}{\del A_i}\frac{\del\chns}{\del \Pi_i}-\frac{\del\chns}{\del \Pi_i}\frac{\del\chns}{\del A_i}\right)\nn\\
&=&0\;,
\eea
trivially because the classical operators commute as 
seen by the right-hand sides of Eqs.~(\ref{eq:itworks0}) and (\ref{eq:itworks}).
Dissipation has been included in this argument as given by the right-hand side of Eq.~(\ref{eq:itworks}). 
As is well known, thermodynamics restores energy conservation by consistently including heat 
(in this case from viscous dissipation) into the definition of what is meant by `energy'.

The algebra for the time derivative of an arbitrary observable is just as simple but it does not vanish---rather 
it leads to the definition of a Poisson bracket. 
The variation of a general classical dissipative observable density ${\cal O}$ 
(with at most a first spatial derivative dependence---the standard case, like for $\chns$) 
is given by
\bea
&&\!\!\!\!\!\!\!\!\del O[\ba,\bb,s(\ba)]=\intx\,\del{\cal O}\left(\ba,\bb,s(\ba)\right)=\nn\\
&&\!\!\!\!\!\!\!\!\!\intx\,
\left[\frac{\pd{\cal O}}{\pd A_i}\delta A_i
+\frac{\pd{\cal O}}{\pd (\pj A_i)}\delta (\pj A_i)
+\frac{\pd{\cal O}}{\pd s}\frac{\del s}{\del A_i}\delta A_i
+\frac{\pd{\cal O}}{\pd \Pi_i}\delta \Pi_i
+\frac{\pd{\cal O}}{\pd (\pj \Pi_i)}\delta (\pj \Pi_i)
\right]\nn\\
&&\!\!\!\!\!\!\!\!\!=\int d^2x\,\hat{n}_j
\left[
\frac{\pd{\cal O}}{\pd (\pj A_i)}\delta A_i
+\frac{\pd{\cal O}}{\pd (\pj \Pi_i)}\delta \Pi_i
\right]+\nn\\
&&\!\!\!\!\!\!\!\!\!+\intx\,
\left\{\left[\frac{\pd{\cal O}}{\pd A_i}
-\pj\left(\frac{\pd{\cal O}}{\pd (\pj A_i)}\right)
+\frac{\pd{\cal O}}{\pd s}\frac{\del s}{\del A_i}\right]\delta A_i
+\left[\frac{\pd{\cal O}}{\pd \Pi_i}
-\pj\left(\frac{\pd{\cal O}}{\pd (\pj \Pi_i)}\right)\right]\delta \Pi_i
\right\}
\eea
which implies the following for observable ${\cal O}$'s functional derivatives 
(dropping surface terms like discussed above since independent variations 
$\del\ba$ and $\del\bb$ vanish at spatial infinity by physical assumption)
\bse
\label{kool}
\bea
\frac{\del{\cal O}}{\del A_i}&=&\fbox{$
\frac{\pd{\cal O}}{\pd A_i}
-\pj\left(\frac{\pd{\cal O}}{\pd (\pj A_i)}\right)
+\frac{\pd{\cal O}}{\pd s}\frac{\del s}{\del A_i}$}\label{kool1}
\;,\\
\frac{\del{\cal O}}{\del \Pi_i}&=&\fbox{$
\frac{\pd{\cal O}}{\pd \Pi_i}
-\pj\left(\frac{\pd{\cal O}}{\pd (\pj \Pi_i)}\right)$}
\;.
\eea
\ese
Allowing the variation to be a time derivative (see discussion below Eq.~(\ref{eq:scon3})), the above becomes
\bea
&&\!\!\!\!\!\!\!\!\frac{dO[\ba,\bb,s(\ba)]}{dt}=\nn\\
&&\!\!\!\!\!\!\!\!\!\intx\,
\left\{\left[\frac{\pd{\cal O}}{\pd A_i}
-\pj\left(\frac{\pd{\cal O}}{\pd (\pj A_i)}\right)
+\frac{\pd{\cal O}}{\pd s}\frac{\del s}{\del A_i}\right]\frac{\pd A_i}{\pd t}
+\left[\frac{\pd{\cal O}}{\pd \Pi_i}
-\pj\left(\frac{\pd{\cal O}}{\pd (\pj \Pi_i)}\right)\right]\frac{\pd \Pi_i}{\pd t}
\right\}\label{eq:truetrue}
\eea
which upon inserting Eqs.~(\ref{kool}) and (\ref{eq:bday3}) becomes 
\bse
\bea
\frac{dO[\ba,\bb,s(\ba)]}{dt}&=&\fbox{$
\intx\,\left[\frac{\del{\cal O}}{\del A_i}\frac{\del\chns}{\del \Pi_i}
-\frac{\del{\cal O}}{\del \Pi_i}\frac{\del\chns}{\del A_i}
\right]$}\label{diss}\\
&\equiv&\intx\,\left[{\cal O},\chns\right]\\
&\equiv&\left[O,\hns\right]\;,
\eea
\ese
where the square bracket with a comma separating the observables (classical differential operators)  
is the definition of a Poisson bracket in field theory.  
In summary, the time derivative of an arbitrary dissipative fluid observable, that does not explicitly depend on time, 
is given by its Poisson bracket with the Navier-Stokes Hamiltonian of Eq.~(\ref{eq:hns}): 
\be\fbox{$
\frac{dO}{dt}=\left[O,\hns\right]$}\label{ohya}
\;.
\ee 
This looks like the standard result and this dissipative canonical Poisson bracket structure 
of Eq.~(\ref{diss}) was the primary motivation of introducing $\ba(\bx,t)$ by Fukagawa and Fujitani \cite{hiroki}---but 
note carefully that the functional derivatives 
$\del{\cal O}/\del A_i$ and $\del\chns/\del A_i$ of this Poisson bracket are dissipative and given by
Eqs.~(\ref{kool1}) and (\ref{eq:itworks}) respectively, where  
 $\del s/\del A_i$
is the same entropy constraint, Eq.~(\ref{eq:000b}), used throughout this paper.

We end by working out the Poisson bracket between the Navier-Stokes Hamiltonian 
coordinate field $\ba(\bx^\pr,t)$ and its conjugate momentum field 
$\bb(\bx^{\pr\pr},t)$ at two different field points at the same time (left implicit after this). 
The canonical result is a delta function and we show  
that it follows here too. 
A Poisson bracket of the dynamical fields of $\hns$ is 
\bea
\left[A_i(\bx^\pr),\pi_{A_j}(\bx^{\pr\pr})\right]&=&\left[A_i(\bx^\pr),\Pi_j(\bx^{\pr\pr})\right]\nn\\
&\stackrel{(\ref{diss})}{=}&\intx\left[\frac{\del A_i(\bx^\pr)}{\del A_k(\bx)}\frac{\del \Pi_j(\bx^{\pr\pr})}{\del \Pi_k(\bx)}
-\frac{\del A_i(\bx^\pr)}{\del \Pi_k(\bx)}\frac{\del \Pi_j(\bx^{\pr\pr})}{\del A_k(\bx)}\right]\nn\\
&=&\intx\left[\del_{ik}\del^3(x-x^\pr)\del_{jk}\del^3(x-x^{\pr\pr})
-0\cdot 0\right]\nn\\
&=&\del_{ij}\del^3(x^\pr-x^{\pr\pr})
\;,
\eea
where the third line follows from the standard rules of functional differentiation. 
The standard result follows here because $\ba$  and $\bb$ themselves do not depend on 
entropy $s(\ba)$, rather just $\chns$ or in general ${\cal O}$ does.

\section{Summary and Discussion}

The Navier-Stokes Hamiltonian was derived from first principles. 
The mass, momentum, and energy conservation equations of a compressible viscous fluid 
were shown to lead to the standard entropy constraint. Viscous dissipation in a fluid 
generates heat and the entropy of the fluid necessarily increases according to the 
local laws of thermodynamics. The entropy constraint became a nonholonomic constraint 
for both the Lagrangian and Hamiltonian of the viscous fluid. 
The Euler-Lagrange and Hamilton equations of a viscous fluid 
were shown to be equivalent to the Navier-Stokes equation with mass and energy conservation. 
The Lagrangian and Hamiltonian of the ideal fluid were also shown 
to give equivalent dynamics to that of the Euler equation with mass and inviscid energy conservation. 
This was done to compare 
the viscous and inviscid theories since whether they are smoothly related or not as the 
viscosity vanishes is still an open problem. The dissipative canonical Poisson bracket structure 
of the viscous fluid with dissipative functional derivatives was derived. This dissipative 
Hamiltonian algebra sets up future work with similarity renormalization group methods applied 
to viscous (and inviscid) fluids so as to resolve the respective dynamical scales 
in a systematic fashion. 
Hamiltonian methods allow convenient approximations through the variational principle and 
renormalization group transformations.

Note that the ideal and viscous fluids have quite different pathlines as given 
by the velocity constraint that came from both procedures:
\bse
\label{billy}
\bea
\bv_E&=&\fbox{$\bn\phi-\frac{\alpha}{\rho}\bn\beta-\frac{\lam}{\rho}\bn s$}=
\bn\pi_\rho-\frac{\pi_\beta}{\rho}\bn\beta-\frac{\pi_s}{\rho}\bn s
\;,\\
\bv_{NS}&=&\fbox{$-\frac{\Pi_i}{\rho}\bn A_i$}=-\frac{\pi_{A_i}\bn A_i}{J(\bn\ba)\,\rho_0(\ba)}\;,
\eea
\ese
from Eqs.~(\ref{eq:v}), (\ref{eq:nsv}) and (\ref{eq:masscon})
 respectively.\footnote{To be clear: everywhere in this paper, 
repeated indices are assumed to be summed over three spatial dimensions. Also, as explained in the body of the paper, 
vector field $\ba$ is just another coordinate like $\bx$, so the fact that there are three scalar potential pairs 
for the viscous fluid, $(A_1,\Pi_1;A_2,\Pi_2;A_3,\Pi_3)$, is directly related to the choice of working in three spatial dimensions. 
The viscous fluid seems to be perfectly coupled to three spatial dimensions.}
There are two points to be made both highlighting the differences between the ideal and viscous fluids 
even though at first sight, these $\bv_E$ and $\bv_{NS}$ decompositions look similar. 
First, for $\bv_E$ note how the $\bn\phi$ term has a plus sign\footnote{Obviously 
$\cle$ of Eq.~(\ref{eq:leuler}) could be defined so that this is a minus 
sign or vice versa on the other two terms of $\bv_E$, but the point is 
that then the conjugate momentum field for one of the $\bv_E$ terms has to have an opposite sign to 
the field in the Lagrangian that it represents. This sign asymmetry 
does not occur for the viscous fluid. There the signs of all three terms can be made to be the same in $\bv_{NS}$ 
and the conjugate momenta of $\chns$ are related by a plus sign to the respective 
field in the Lagrangian: $\pi_{A_i}=\,\Pi_i$.}  
and also this first term does not have any density dependence. 
$\phi$ is of course the standard velocity potential of fluid dynamics and $\bv_E$ is the 
so-called Clebsch decomposition of the velocity field 
with Gauss potential \cite{jackiw} pairs ($\alpha$,$\beta$) and ($\lam$,$s$). Interestingly, in order to obtain the 
Navier-Stokes Hamiltonian, one was required to introduce coordinate transformation $\ba(\bx,t)$ 
in order to handle the dissipation constraint properly, and with its conjugate momentum fields, $\bb(\bx,t)$,  
this already gave enough degrees of freedom ($6$ scalar potentials in $3d$) and 
so a velocity potential was not required. The ideal and viscous fluids are very different and 
this may be one of the origins of this difference. 
Second, $\bv_E$ and $\bv_{NS}$ being different is even more readily apparent 
if we recall that $\rho$ is a dynamical field for the ideal fluid (satisfying the standard continuity equation), 
whereas for the viscous fluid, $\rho=J\rho_0$ is a constraint---which because of the Jacobian factor 
is quite complex: recall Eq.~(\ref{eq:6terms}) for explicit expressions for $J(\bn\ba)$. Also note 
that since $J(\bn\ba)$ with derivatives of fields is in the denominator, it is a nonlocal operator in $\hns$. 
These differences should not come as a huge surprise since 
the ideal and viscous fluid problems apparently do not map smoothly onto each other with   
the zero viscosity and infinitesimal viscosity fluids not being limits of the same theory. 
Perhaps these variational principle forms of the equations will help make this clearer.

\appendix

\section{Classical Similarity Renormalization Group}
\label{app:csrg}

The story of the classical similarity renormalization group (SRG) begins with 
the construction of Wilson's renormalization group around 1965 \cite{rg1,rg2,rg3}  
to integrate the scales of a field theory one momentum shell at a time in 
order to solve the problem with less degrees of freedom for more practical numerical implementations. 
Then---much later starting c.~Autumn $1992$---in order to handle the small-energy denominators that arise 
in hamiltonian field theory perturbation theory, the \emph{similarity renormalization group} 
was developed at Ohio State by G{\l}azek and Wilson \cite{srg}. 
Ohio State was working on the QCD problem of the theory of quarks and gluons.  
Independently, around the same time, the SRG was also developed 
for condensed matter physics by Wegner \cite{floweqn}  
under the name \emph{flow equation}. The SRG was further developed 
for low-energy nuclear physics starting around 2007 
by Furnstahl, Perry, and colleagues at Ohio State \cite{srgosu}. These are all 
quantum applications of the SRG. Below, the SRG is extended to classical physics. 
Eventually both problems may feedback on each other helping one to solve 
the other as a function of scale.

The definition of the classical SRG starts with the quantum flow equation for a 
Hamiltonian of Wegner form (here and only here, `s' is the running ``size'' scale of the flow; 
everywhere else in this paper,
`s' is the specific entropy field):
\be
\frac{dH}{ds}=\left[\left[H_0,H\right]_{{\rm q.c.}},H\right]_{{\rm q.c.}}\;,
\label{eq:qc}
\ee
where $H_0$ is an initial Hamiltonian of interest: $H=H_0+V$, chosen to get the equation to close on itself 
(at least approximately). The square brackets here 
are quantum commutators (as labeled by `q.c.'; just below without the `q.c.', 
a Poisson bracket of classical mechanics \cite{goldstein} is implied), 
but now we make them Poisson brackets through 
the correspondence principle of Bohr, Dirac, and co-founders \cite{dirac1,sakurai}. 
Thus the classical SRG equation starts with  
\beaa
\frac{dH}{ds}=(i\hbar)^2\left[\left[H_0,H\right],H\right]~~~~
\Rightarrow~~~~\frac{dH}{d\lam}=-\left[\left[H_0,H\right],H\right]=\left[H,\left[H_0,H\right]\right]
\;,
\eeaa
where $\hbar$ has been absorbed into the 
definition of the classical size scale of the flow\footnote{ 
Notation warning: at Eq.~(\ref{eq:leuler}) and after, $\lam$ is a lagrange multiplier field in the 
ideal fluid section of the body of the paper. In this appendix, $\lam$ is the classical size scale.}: 
$\lam=s\hbar^2$. The dimensions 
of $\lam$ are  
$({\rm time})^2$. $\lam$ can be thought of as a larger and larger size resolution of the 
system: as $\lam$ is run from small to large values, it is like zooming out with a microscope. 
This flow equation (synonymous here with ``classical SRG equation'') 
generates the effective interactions of a Hamiltonian 
as the dynamics of the system are ``zoomed-out'' to the long-distance scales of interest. 
The final formula for the classical SRG equation in terms of its scale generator $\eta_\lam$ is
\be
\fbox{$\frac{dH}{d\lam}=\left[H,\left[H_0,H\right]\right]=\left[H,\eta_\lam\right]$}
\;,
\label{eq:cSRG}
\ee
with $H=H_0+V$.
Comparing with dynamical evolution, Eq.~(\ref{ohya}), with $dt$ on the \emph{left} and time-translation generator $\hns$ on the 
\emph{right}, here with scale evolution, we have $d\lam$ on the \emph{left} and scale generator $\eta_\lam$ on the \emph{right}. 
The definitions appear to be consistent. Notation note: 
$\eta$ itself in this work is shear viscosity; $\eta_\lam$ which depends on scale $\lam$, is the 
scale generator. Eq.~(\ref{eq:cSRG})
looks the same as the quantum flow equation, Eq.~(\ref{eq:qc}), except for an overall 
minus sign difference and now the square brackets imply  ``Poisson bracket''. 
In order to check this sign and to introduce the classical SRG further, 
now we look at some simple examples: 
first the fixed-source problem which is easy to work out in the classical and quantum arenas, then 
some toy problems to warm up for the viscous fluid, and finally we close with an example flow equation of  
$\hns$ and discuss its similarities with the simple toy problems.

An example classical fixed-source problem is a unit-mass harmonic oscillator in a near-earth gravitational field. 
Here we integrate out the gravitational field interaction with the classical SRG.
The math is illustrative of the classical SRG technique and also maps 
term by term to the quantum problem \cite{bdjfs}; thus we see that the signs of 
Eq.~(\ref{eq:qc}) and Eq.~(\ref{eq:cSRG}) are both correct (even though they are opposite). 
A classical fixed-source Hamiltonian in one spatial dimension is given by 
\beaa
H=\frac{p^2}{2}+\frac{x^2}{2}+g\,x
\;,
\eeaa
where g is a constant (e.g.\ $9.8\,m/s^2$). 
Choosing $H_0=\frac{p^2}{2}+\frac{x^2}{2}$, 
the scale generator for this problem, from Eq.~(\ref{eq:cSRG}), becomes
\beaa
\eta_\lam&=&\left[H_0,H\right]=\left[\frac{p^2}{2}+\frac{x^2}{2},\,g\,x\right]=-p\,g\;,
\eeaa
where recall a Poisson bracket is defined by \cite{goldstein}
\beaa
\left[A,B\right]=\frac{\pd A}{\pd x}\frac{\pd B}{\pd p}-\frac{\pd A}{\pd p}\frac{\pd B}{\pd x}
\;.
\eeaa
Actually, this is a canonical Poisson bracket, and one of the main points of 
the form of Fukagawa and Fujitani's 
Navier-Stokes Hamiltonian \cite{hiroki} is that this canonical 
Poisson bracket structure can be maintained for a 
viscous fluid too (albeit with dissipative functional derivatives). 
We have made a specific choice of $H_0=\frac{p^2}{2}+\frac{x^2}{2}$ here; 
this corresponds to the usual choice made for the 
analogous quantum problem \cite{bdjfs}. Part of the art of defining 
the SRG is to pick a reasonable $H_0$. This issue is discussed further with toy examples just below.

Continuing with the classical fixed source, its flow equation becomes
\beaa
\frac{dH}{d\lam}&=&\left[H,\eta_\lam\right]=\left[\frac{p^2}{2}+\frac{x^2}{2}+g\,x,-p\,g\right]=
(x+g)(-g)=-g^2-g\,x\;.
\eeaa
In order for this equation to close on itself we have to add a constant ($x$, $p$ indep.) self-energy to the Hamiltonian 
and the coupling has to run with $\lam$ as well:
\beaa
\Rightarrow H \longrightarrow H_\lam = \Sigma_\lam+\frac{p^2}{2}+\frac{x^2}{2}+g_\lam\,x\;.
\eeaa
Given $\frac{dH}{d\lam}$ derived above, these two ``couplings'' of this Hamiltonian must run according to
\beaa
\frac{dg_\lam}{d\lam}=-g_\lam\;,~~~~
\frac{d\Sigma_\lam}{d\lam}=-g_\lam^2\;.
\eeaa
These flow equations for the couplings are easy to solve and we have
\beaa
g_\lam=g_0\,e^{-\lam}\;,~~~~
\Sigma_\lam=\Sigma_{\lam_0}-\frac{g_0^2}{2}\left[e^{-2\lam_0}-e^{-2\lam}\right]
\;.
\eeaa
This result for the running couplings corresponds precisely with the quantum fixed-source problem \cite{bdjfs} and 
we see that as $\lam\rightarrow\infty$ (i.e.\ the microscope is ``zoomed-out''), 
the gravity interaction (here $g_\lam\,x$) is integrated out, and 
the self-energy $\Sigma_\lam$ is left 
dressed with a background field like the Yukawa meson-cloud left over with the quantum 
fixed-source problem \cite{bdjfs}.
The parallelism between the quantum and classical problems here is quite striking 
showing the correct sign identification for the above classical SRG equation.

\begin{quote}\emph{
The classical SRG is a new technique, so we illustrate it further now 
with three toy examples and then close with a flow 
equation of $\hns$. Part of the question here is what is the best $H_0$ to use for the particular 
problem one is trying to solve. This is still an open question. The classical SRG may help to 
answer this question (even the quantum one).}
\end{quote}

Next we show some results of the classical SRG flow for three general interactions. The first 
two examples are the same Hamiltonian, but with different choices for $H_0$, and then 
the final example is a toy made to be similar to $\hns$, albeit without its nonlocalities 
(so this is an incompressible toy). 
Nevertheless, as shown at the end of this section, this third toy example has nice similarities with the 
full flow equation of $\hns$. 

For these three examples, $n$ is a positive integer: $n>0$. 
The fixed-source problem just discussed is $n=1$; $n=4$ is the quartic oscillator interaction, etc. 
For all of these examples, $H\equiv H_0 + V$, where $V$ is an interaction potential. The three examples are 
shown in the following table. 
\begin{center}
    \begin{tabular}{ | p{4cm} | p{2.5cm} | p{2.5cm} |}
    \hline
		example & $H_0$ & $V$ \\ \hline
		1. $H_0$ free&
		$p^2/2$ &
		$x^2/2+g\,x^n/n$ \\ \hline
		2. $H_0$ oscillator&
		$p^2/2+x^2/2$ & 
		$g\,x^n/n$ \\ \hline
		3. $\hns$ toy&
    $p^2x^2/2$ &
		$g\,e^{-kx}$ \\  \hline
    \end{tabular}
\end{center}
Define force $F\equiv-\pd V/\pd x$, and keep in mind that $V$ changes for each of these examples. 
We see a nice pattern appearing: the SRG evolution for these three examples is given by the following table.
\begin{center}
    \begin{tabular}{ | p{4cm} | p{1.5cm} | p{7cm} |}
    \hline 
		example & $\eta_\lam$ & $dH/d\lam$ \\ \hline
		1. $H_0$ free&
		$p F$ &
		$2H_0\,\pd^2V/\pd x^2-F^2$ \\ \hline
		2. $H_0$ oscillator&
		$p F$& 
		$-x\pd V/\pd x+p^2\pd^2V/\pd x^2-F^2$ \\ \hline
		3. $\hns$ toy&
    $p F x^2$ &
		$2H_0(x\pd V/\pd x+x^2\pd^2V/\pd x^2)-F^2x^2$ \\  \hline
    \end{tabular}
\end{center}
Two things in particular to note here: (1) the $\hns$ toy with ``quartic $H_0$'' 
is not all that different from the other two and (2) these patterns in terms of 
$F\equiv -\pd V/\pd x$ were noticed because we were using the classical SRG with 
its Poisson brackets---the $\pd/\pd x$ terms naturally appear with the classical SRG.

Now we write the corresponding flow equations for the full $\hns$. 
For $\hns$ we define (this is not the only way to do it, but it appears to be natural 
and is what we start with; this may be a good choice for a thermal fluid) 
free and interacting hamiltonian densities
\beaa
\ch_0\left(\ba,\bb\right)&\equiv&\frac{(\Pi_i\bn A_i)^2}{2\,J(\bn\ba)\,\rho_0(\ba)}\;,\\
\cv\left(\ba,\bb,s(\ba)\right)&\equiv&\cu(J(\bn\ba)\,\rho_0(\ba),s(\ba))\;.
\eeaa
With $\hns\equiv H_0+V$, $H_0=\intx\,\ch_0$ and $V=\intx\,\cv$, the classical SRG equation becomes
\beaa
\frac{d\hns}{d\lam}=\left[\hns,\eta_\lam\right]=\left[\hns,\left[H_0,V\right]\right]\;.
\eeaa
After some detailed but straightforward algebra with dissipative canonical 
Poisson brackets, as 
defined in the body of this paper, the scale generator of $\hns$ becomes 
\bea
\eta_\lam\equiv\intx\,\ceta_\lam=\intx\,\left[\ch_0,\cv\right]=\intx\,v_i\pj\sig_{ij}\;,
\label{eq:woweta}
\eea
where $\sig_{ij}$ is the full stress tensor of a viscous fluid, Eq.~(\ref{eq:3}). 
Note that this is just a ``velocity times force'' term as is the $\eta_\lam=-p\,g$ of the 
fixed source unit-mass example above. 
Actually, we see that this result for $\eta_\lam$ of Eq.~(\ref{eq:woweta}) is  
the precise term of the right-hand side of the kinetic energy conservation equation: 
the work done on the system (per unit time per unit volume) 
by the pressure and viscous forces of the fluid. 
This provides a nice physical interpretation for scale generator $\eta_\lam$: 
\emph{the power required to integrate out the interaction of interest}.

Now for $d\hns/d\lam$. This too is involved, but the algebra is straightforward and at the same level 
as deriving the hamilton equations of $\hns$ in the body of this paper---cf. Eq.~(\ref{100}). 
The SRG flow of $\hns$ for the $H_0$ choice shown above is given by
\beaa
\frac{d\hns}{d\lam}=\intx\,\left\{\left[\ch_0,\ceta_\lam\right]+\left[\cv,\ceta_\lam\right]\right\}\;,
\eeaa
with result  
\beaa
\left[\ch_0,\ceta_\lam\right]&=&v_\ell\frac{\del s}{\del A_k}\left\{
2\rho T (\pj A_k)(\partial_\ell v_j)+\pj[\rho T v_j (\partial_\ell A_k)]\right\}\;,\\
\left[\cv,\ceta_\lam\right]&=&-\frac{(\pj\sig_{ij})^2}{\rho}\;.
\eeaa
Recall Eq.~(\ref{eq:000b}) for the entropy constraint. 
Before discussing these two results, we need one further result to relate $d\hns/d\lam$ to the toy model results:
\beaa
\frac{\del\cv}{\del A_i}\equiv\frac{\del\cu}{\del A_i}=-(\pj h)\rho 
\frac{\pd x_j}{\pd A_i}+\rho T\frac{\del s}{\del A_i}
\;,
\eeaa
where $\cu$ is the internal energy density of $\hns$, and this 
came straight from the hamilton equations of $\hns$ with full 
algebra shown in the body of this paper---cf. Eq.~(\ref{100}). Also, note an implicit way of writing the free Hamiltonian
here is $\ch_0=\rho \bv^2/2$. Thus altogether, corresponding to the above table, we can identify
\beaa
-F^2&\longrightarrow&-\frac{(\pj\sig_{ij})^2}{\rho}\\
H_0\,x\,\pd V/\pd x&\longrightarrow&\sim\left[\ch_0,\ceta_\lam\right]
\eeaa
at a qualitative level (with classical mechanics to field theory correspondences $x\rightarrow\ba(\bx,t)$ and 
$\pd V/\pd x\rightarrow\frac{\del\cv}{\del A_i}$), 
and we see how the results of the classical mechanics toy models can be useful 
in interpreting field theory results.
Note that the $\pd^2V/\pd x^2$ term of the toy models is gone 
in this example of the flow of $\hns$ because the nonholonomic constraint relates 
$\del s\sim\del A_i$, but there is no $\del^2s$ term.

\section{Symmetries of a nonrelativistic viscous fluid}
\label{app:sym}

A nonrelativistic viscous fluid has ten galilean group generators including the 
Hamiltonian, momentum density, angular momentum density, and boosts: 
\{$\hns$, $\rho \bv$, $\rho\bx\times\bv$, $\rho\bx$\} \cite{sudarshan}. These generators 
correspond with time-translations, 
space-translations, rotations, and galilean boosts respectively. The first three are symmetries 
(with the respective Poisson bracket with the Hamiltonian vanishing), the galilean boosts are not 
symmetries but they have a simple interpretation as shown next. 
Keep in mind that
these densities are written in ``implicit form'': really everything needs 
to be expressed in terms of the Navier-Stokes Hamiltonian coordinate fields $\ba(\bx,t)$ and $\bb(\bx,t)$. 
For example, explicitly we have
\beaa
\rho\bv&=&-\Pi_i\bn A_i\;,\\
\rho&=&\rho_0(\ba)J(\bn\ba)\;,
\eeaa
where J is the Jacobian determinant---cf.\ Eq.~(\ref{eq:6terms})---given
by $J=~\eps_{ijk}\,(\pdi A_1)(\pj A_2)(\pk A_3)$.

The galilean algebra was checked for all ten of these generators of a nonrelativistic viscous fluid 
based on $\hns$. 
The Poisson brackets with $\hns$ itself are special because they are the time derivative 
of the respective in general dissipative generator:
\beaa
\frac{dO[\ba,\bb,s(\ba)]}{dt}=
\intx\,\left[\frac{\del{\cal O}}{\del A_i}\frac{\del\chns}{\del \Pi_i}
-\frac{\del{\cal O}}{\del \Pi_i}\frac{\del\chns}{\del A_i}
\right]
\equiv\intx\,\left[{\cal O},\chns\right]
\equiv\left[O,\hns\right]\;.
\eeaa
So, when this vanishes, the respective generator corresponds to a symmetry of the theory.
For the galilean group there are three symmetries (along with mass conservation) 
for a nonrelativistic viscous fluid as shown 
by the following table.
\vspace{-12pt}
\begin{center}
    \begin{tabular}{ | p{8.25cm} |  p{4.5cm} | p{4.25cm} |}
    \hline 
		\textbf{time derivative} & \textbf{conserved quantity} & \textbf{symmetry} \\ \hline
		$d\hns/dt=[\hns,\hns]=0$&
		 fluid energy & 
		time translation\\ \hline
		$d(\intx\,\rho\bv)/dt=\intx\,[\rho\bv,\chns]=0$&
		 fluid momentum & 
		translational invariance\\ \hline
		$d(\intx\,\rho\bx\times\bv)/dt=\intx\,[\rho\bx\times\bv,\chns]=0$&
		 fluid angular momentum & 
		rotational invariance\\ \hline
		$d(\intx\,\rho)/dt=\intx\,[\rho,\chns]=0$&
		 fluid mass & 
		continuity\\ \hline
    \end{tabular}
\end{center}
\vspace{6pt}
This last entry, total mass conservation, is an extension to the ten galilean group generators, 
and its conservation is a result of the continuity of the fluid. 
The rest of the galilean algebra was checked and 
in summary, the nonzero Poisson brackets are 
(the boost and physical momentum generators, $\vect{B}$ and $\vect{P}$, 
are defined just after this block of equations)
\beaa
\frac{dB_i}{dt}&=&\left[B_i,\hns\right]=P_i\;,\\
\left[B_i,P_j\right]=M\del_{ij}\;,~~
\left[J_i,J_j\right]&=&\varepsilon_{ijk}J_k\;,~~
\left[J_i,P_j\right]=\varepsilon_{ijk}P_k\;,~~
\left[J_i,B_j\right]=\varepsilon_{ijk}B_k\;,
\eeaa
where $\varepsilon_{ijk}$ is the Levi-Civita symbol with $\varepsilon_{123}= 1$; 
$\del_{ij}$ is the Kronecker delta; and the boost generator, 
physical momentum, angular momentum, 
and total mass have been defined by
\beaa
\vect{B}\equiv\intx\,\rho\bx\;,~~
\vect{P}\equiv\intx\,\rho\bv\;,~~
\vect{J}\equiv\intx\,\rho\bx\times\bv\;,~~
M\equiv\intx\,\rho\;.
\eeaa
Along with $\hns$, these are the standard galilean group generators showing that $\vect{J}$ is an angular momentum 
and that $\vect{B}$ and $\vect{P}$ are vectors according to their rotation properties. 
\begin{quote}\emph{
The point is that 
$\hns$ is complicated by a nonholonomic entropy constraint, but this can be dealt with in a 
straightforward way with a dissipative functional derivative---cf. Eq.~(\ref{eq:itworks})---and 
all the rest of the algebra follows the canonical Poisson bracket structure. That was 
the primary motivation of introducing $\ba(\bx,t)$ in $\hns$ by Fukagawa and Fujitani} \cite{hiroki}.
\end{quote}
Given mass conservation of the last row of the table above, $dM/dt=0$, 
this $d\vect{B}/dt$ term can be rewritten in an intuitive way after 
defining a center-of-mass vector:
\beaa
\crr\equiv\frac{\vect{B}}{M}=\frac{\intx\,\rho\,\bx}{\intx\,\rho}\;.
\eeaa
\begin{quote}\emph{
Remember, $\rho=\rho_0J\sim\rho_0(\ba)\sum\pm(\pd A_i/\pd x)(\pd A_j/\pd y)(\pd A_k/\pd z)$, 
but it all works. }
\end{quote}
So, to close the discussion, given mass and momentum conservation,
we see the time derivative 
of the galilean boost generator for a Navier-Stokes fluid is nothing more than
\beaa
\frac{d\vect{B}}{dt}=M\frac{d\crr}{dt}=\vect{P}\;,
\eeaa
which since total physical momentum $\vect{P}$ is constant, is easy to integrate 
and we are led to the standard result for a nonrelativistic viscous fluid:
\beaa
\crr=\crr(0)+\vect{P}\,t/M
\;,
\eeaa
with the total center-of-mass of the fluid obeying free motion.
This all followed from $\hns$ of Eq.~(\ref{eq:hns}). 
It is a complicated in general nonlocal operator, but this paper showed that it can be dealt 
with in a straightforward and intuitive way through its dissipative canonical Poisson bracket structure.

\begin{acknowledgments}
The author would like to thank Hiroki Fukagawa, Dick Furnstahl, and  
Robert Perry for enlightening discussions while these ideas were being put together. 
This work was supported in part by the University of Washington Applied Physics Laboratory  
under Project \mbox{No.~APL-UW--900088}.
\end{acknowledgments}

\end{document}